\documentclass[fleqn,usenatbib]{mnras}

\usepackage{ae,aecompl}
\usepackage{graphicx}
\usepackage[dvipsnames]{xcolor}
\usepackage{amsmath}
\usepackage{amssymb}
\usepackage{savesym}
\savesymbol{tablenum}
\usepackage{makecell}
\usepackage{siunitx}
\usepackage{xspace}
\usepackage{listings}
\usepackage{soul}
\usepackage{float}
\restoresymbol{SIX}{tablenum}

\usepackage{comment}

\usepackage[T1]{fontenc}
\usepackage{fontawesome}
\definecolor{linkcolor}{HTML}{2AAD2E}

\usepackage{newtxtext,newtxmath}

\newcommand{\fr}{\texttt{frank}\xspace}
\newcommand{\cl}{\texttt{CLEAN}\xspace} 
\newcommand{\ws}{$w_{\rm smooth}$\xspace}
\newcommand{\al}{$\alpha$\xspace}
\newcommand{\ml}{M$\lambda$\xspace}
\newcommand{\kl}{k$\lambda$\xspace}
\newcommand{\bef}{$B_{\rm 80,\ frank}$\xspace}
\newcommand{\bec}{$B_{\rm 80,\ CLEAN\ image}$\xspace}
\newcommand{\bem}{$B_{\rm 80,\ CLEAN\ model}$\xspace}
\newcommand{\bed}{$B_{\rm data,\ expected}$\xspace}

\title[A super-resolution analysis of DSHARP]{A super-resolution analysis of the DSHARP survey: Substructure is common in the inner 30 au}

\author[Jennings, Booth, Tazzari, Clarke, Rosotti]{
Jeff Jennings,$^{1}$\thanks{E-mail: jmj51@ast.cam.ac.uk} Richard A. Booth,$^{2}$ Marco Tazzari,$^1$ Cathie J. Clarke,$^1$ and \newauthor Giovanni P. Rosotti$^{3,4}$
\\
$^1$Institute of Astronomy, University of Cambridge, Madingley Road, Cambridge CB3 0HA, UK\\
$^2$Astrophysics Group, Imperial College London, Prince Consort Road, London SW7 2AZ, UK \\
$^3$Leiden Observatory, University of Leiden, P.O. Box 9500, Leiden NL-2300 RA, Netherlands \\
$^4$School of Physics and Astronomy, University of Leicester, Leicester LE1 7RH, UK
}

\date{Accepted XXX. Received YYY; in original form ZZZ}

\pubyear{2021}

\begin{document}
\label{firstpage}
\pagerange{\pageref{firstpage}--\pageref{lastpage}}
\maketitle

\begin{abstract}
The DSHARP survey evidenced the ubiquity of substructure in the mm dust distribution of large, bright protoplanetary discs. Intriguingly, these datasets have yet higher resolution information that is not recovered in a \cl image. We first show that the intrinsic performance of the \cl algorithm is resolution-limited. Then analyzing all $20$ DSHARP sources using the 1D, super-resolution code Frankenstein (\fr), we accurately fit the 1D visibilities to a mean factor of $4.3$ longer baseline than the Fourier transform of the \cl images and a factor of $3.0$ longer baseline than the transform of the \cl component models. This yields a higher resolution brightness profile for each source, identifying new substructure interior to $30$ au in multiple discs; resolving known gaps to be deeper, wider, and more structured; and known rings to be narrower and brighter. Across the survey, high contrast gaps are an average $14\%$ wider and $44\%$ deeper in the \fr profiles relative to \cl, and high contrast rings are an average $26\%$ narrower. Categorizing the \fr brightness profiles into trends, we find that the relative scarcity of features interior to $30$ au in the survey's \cl images is an artifact of resolving power, rather than an intrinsic rarity of inner disc (or compact disc) substructure. Finally the rings in the \fr profiles are narrower than the previously inferred deconvolved widths, indicating smaller $\alpha / {\rm St}$ ratios in the local gas disc.
\end{abstract}


\section{Physical and methodological contexts}
\label{sec:intro}
Interferometric observations of the dust and gas components of protoplanetary discs provide the highest resolution information available on the structure of these sources, which in turn traces the planetary companions and physical mechanisms responsible.
At the highest angular resolutions achieved to date in the sub-mm -- mm with the Atacama Large Millimeter Array (ALMA; beam widths of $\approx 25 - 75$ mas corresponding to $\approx$1 -- 10 au)\footnote{Notation: we use $\approx$ to mean \lq{}approximately equal to\rq{} and $\sim$ to mean \lq{}of order.\rq{}}, studies prior to DSHARP first identified, characterized and analyzed an abundance of dust substructure in individual systems \citep{Partnership2015a, Andrews2016, Clarke2018}. The DSHARP survey \citep{Andrews2018, 2018ApJ...869L..42H} then confirmed that annular gaps and rings are ubiquitous in the dust of large, bright discs around single stars. The survey additionally found a nonnegligible occurrence rate of nonaxisymmetric dust substructure in the form of spirals arms \citep{2018ApJ...869L..43H, 2018ApJ...869L..44K} and azimuthally localized brightness arcs \citep{2018ApJ...869L..49I, 2018ApJ...869L..50P}. Studies since DSHARP for individual systems at comparable resolutions have upheld the high occurrence rate of dust substructure \citep[e.g.,][]{2018ApJ...868L...5K, Keppler_2019, 2019NatAs.tmp..419P, 2019arXiv190205143P}.

Analyses of interferometric datasets in the protoplanetary disc community, including in the DSHARP survey, generally rely on images generated with the \cl deconvolution technique \citep{1974A&AS...15..417H, 1980A&A....89..377C, 2008ISTSP...2..793C}. In the reconstruction of a model image from an interferometric measurement, a fundamental challenge is accounting for unsampled spatial frequencies (baselines). A direct Fourier transform of the visibilities at sampled baselines (i.e., an assumption of zero power on unsampled baselines) yields a \lq{}dirty image,\rq{} i.e., the sky brightness convolved with the instrument's point spread function (PSF; \lq{}dirty beam\rq{}). This convolution introduces artifacts into the dirty image due to the PSF's sidelobe structure, and the \cl algorithm is a nonlinear, procedural approach to remove these artifacts (deconvolution). To do this, \cl begins with a \lq{}residual image\rq{} that is equal to the dirty image, then iteratively: finds the peak flux in the residual image, adds a corresponding \lq{}component\rq{} (in the simplest case, a Dirac $\delta$ function) to the \lq{}\cl model\rq{} (an image composed only of the \cl components), and subtracts this component convolved with the dirty beam from the residual image. At the end of this iteration, the \cl model is convolved with the \lq{}\cl beam\rq{} (a Gaussian fit to the primary lobe of the dirty beam), and the final state of the residual image is added to this to form the \lq{}\cl image\rq{} commonly presented as the astronomical observation.

While the \cl algorithm is the standard and highly successful technique used across much of radio interferometry, the procedure imposes artificial resolution loss in the image reconstruction, primarily from convolution of the \cl model with the \cl beam. This causes all features in the \cl image, regardless of their scale, to be smeared in resolution over the size of the beam. For the 1D (radial) brightness profile of a source, convolution induces a reduction in amplitude of all disc features, an overestimate of ring widths, and an underestimate of gap widths.

As we will demonstrate, \lq{}super-resolution\rq{} imaging techniques can overcome the resolution limits of the \cl algorithm.\footnote{By \lq{}super-resolution\rq{} we mean an achieved fit resolution higher than the achieved \cl resolution, which we will quantify as distinct from the \cl beam width.} These methods thus have the capacity to provide new insights into a source's substructure {\it from existing datasets}, better informing physical inference and follow-up observing strategies.
While super-resolution approaches have been applied to individual DSHARP discs, namely parametric visibility fitting in \citealt{Guzman2018}, \citealt{2018ApJ...869L..49I} and \citealt{2018ApJ...869L..50P}, no study has yet examined the entire DSHARP sample.

Super-resolution fitting techniques used in (and in some cases tailored to) the protoplanetary disc field can be divided into image plane and Fourier domain approaches. Image plane procedures include the maximum entropy method \citep{1978Natur.272..686G, 1986ARA&A..24..127N, 2006ApJS..162..401S, 2006ApJ...639..951C, 2013Natur.493..191C, 2016ApJ...829...11C} and sparse modeling \citep{10.1093/pasj/psu070, 2017AJ....153..159A, 2018ApJ...858...56K, 2019ASPC..523..143N}, with the broad class of regularized maximum likelihood techniques being actively used in Very Long Baseline Interferometry \citep[][and references therein]{2019ApJ...875L...4E} and for application to protoplanetary discs \citep{ian_czekala_2021_4939048}. Fourier domain approaches include fitting the visibilities parametrically \citep{PERKINS201573, 2018MNRAS.476.4527T} and nonparametrically \citep{frank_method}.

In this work we characterize substructure at super-resolution scales in all $20$ DSHARP discs using the 1D code \fr \citep{frank_method}, which reconstructs a disc's brightness profile by nonparametrically fitting the azimuthally averaged visibility distribution.\footnote{The code is available at \href{https://github.com/discsim/frank}{\color{linkcolor}{https://github.com/discsim/frank}}. All \fr fits in this work are available at \href{https://zenodo.org/record/5587841}{\color{linkcolor}{https://zenodo.org/record/5587841}}.} Sec.~\ref{sec:model} summarizes the \fr modeling approach and its limitations.
Sec.~\ref{sec:methodologies} then examines the resolution limitations of \cl images and models in real and visibility space (Sec.~\ref{sec:convolution_theory}), compares the accuracy of brightness profiles extracted from the \cl images and models with the \fr visibility fits for the DSHARP sources (Sec.~\ref{sec:clean_frank_accuracy}), and summarizes the principles of comparing \fr to \cl (Sec.~\ref{sec:frank_clean_context}).
In Sec.~\ref{sec:result} we present the super-resolution \fr fits for each DSHARP source, then group the \fr brightness profiles by previously unidentified substructure trends in Sec.~\ref{sec:analysis}. We further use the super-resolution fits to identify a geometric viewing effect that can imprint on disc images. Sec.~\ref{sec:conclusions} summarizes our findings and briefly places them in the context of super-resolution substructure that may be present in other protoplanetary disc datasets, as well as the physical inference this can inform.

\section{Model}
\label{sec:model}
For a full description of the model framework in \fr, see \citet{frank_method}. Here we briefly and qualitatively summarize the approach. \fr reconstructs the azimuthally averaged brightness profile of a source as a function of disc radius by directly fitting the real component of the deprojected, unbinned visibilities as a function of baseline.\footnote{We will use the disc geometries and phase centers in \citet{2018ApJ...869L..42H} to deproject the DSHARP datasets.
Those values were determined in the image plane by either fitting ellipses to individual annular rings or fitting a 2D Gaussian to the image. Across all datasets, we have tested both fitting a 2D Gaussian to the visibilities and fitting the visibilities nonparametrically to determine the geometry and phase center. In general we have found close agreement with the published values and so default to those.} 
The brightness profile is determined nonparametrically by fitting the visibilities with a Fourier-Bessel series, which is linked to the real space profile by a discrete Hankel transform \citep{2015JOSAA..32..611B}.
The Fourier transform of a circle has a Bessel function kernel, making the discrete Hankel transform ({\it DHT}) a natural basis for circular (at least to zeroth order) protoplanetary discs. A Gaussian process regularizes the fit, with the covariance matrix nonparametrically learned from the visibilities under the assumption that this matrix is diagonal in Fourier space. The free parameters (diagonal elements) of the matrix correspond to the power spectrum of the reconstructed brightness profile. The approach is largely built on that in \citet{2013PhRvE..87c2136O}.

The model has five free parameters; variation in reasonable choices for three of these (the outer radius and number of points used in the fit, and the floor value for the power spectral mode amplitudes) has a trivial effect on the recovered profile. Of the remaining two, \al sets the signal-to-noise ({\it SNR}) threshold at which the model stops fitting the data, with a larger \al resulting in a higher SNR threshold. The choice of \al effectively corresponds to a maximum baseline beyond which the model does not attempt to fit the visibilities. 
This is relevant for the DSHARP datasets, as they all become noise-dominated typically at $\gtrsim 5$ \ml, while the maximum baseline is $\approx 10$ \ml. {\it In practice we manually choose an \al value to fit out to the baseline at which the binned visibility SNR begins to oscillate about SNR~=~1} (due to the {\it uv}~sampling becoming highly sparse). The SNR is assessed with $20$~\kl bins of the real component of the visibilities, using ${\rm SNR} = \mu^2 / \sigma^2$, where $\mu$ is the mean visibility amplitude in each bin and $\sigma$ the standard deviation. Pushing the fit out to these long baselines always comes at the cost of fitting some noise, which imprints on the brightness profile as rapid oscillations, usually with very low amplitude (typically $<1\%$ of the profile's peak brightness; as an example, see the fit residuals in Fig.~8 of \citealt{frank_method}).
To suppress these noisy oscillations, the remaining free parameter \ws varies the spatial frequency scale over which the visibility SNR is averaged when building the power spectrum. A nonzero \ws prevents regions of artificially steep gradient in the power spectrum that are due to undersampled baselines. 

For the DSHARP datasets, we use \al and \ws values within the ranges $1.01 \leq \alpha \leq 1.50$ and $10^{-4} \leq w_{\rm smooth} \leq 10^{-1}$, tailoring choices to the unique visibility distribution and noise properties of each dataset. We favor the smaller values within these ranges in order to reduce the constraint placed by the Gaussian process prior on the brightness profile reconstruction.

To fit each dataset, we download the self-calibrated and multi-configuration combined continuum measurement sets from 
\href{https://bulk.cv.nrao.edu/almadata/lp/DSHARP}{https://bulk.cv.nrao.edu/almadata/lp/DSHARP}. Before extracting the visibilities using the \texttt{export\_uvtable} function of the \texttt{uvplot} package \citep{uvplot_mtazzari}, we apply channel averaging (to obtain 1 channel per spectral window) and time averaging (30 sec) to all spectral windows in the original MS table. The \fr fit takes $\lesssim 1$ min for each resulting visibility distribution. 

To generate images of the \fr residual visibilities in this work, we produce measurement sets from the \fr residual UV tables, then use the \texttt{tclean} scripts from the DSHARP website to image. These  scripts yield \cl beams that are often larger than those in the \texttt{.fits} files on the website, though only by $1 - 2$ mas along either axis. The only exception is HD~143006, where the \cl beam is $36 \times 53$ mas in the \texttt{.fits} file, while the \texttt{tclean} script yields $47 \times 48$ mas (this may be due to slightly different versions of {\tt CASA} used). For consistency with the imaged \fr residuals, we will therefore show \cl images generated by applying the published \texttt{tclean} scripts to the published measurement sets, rather than showing the published \texttt{.fits} images.

\subsection{Point source-corrected fits}
\label{sec:point_source_fits}
Eleven of the $20$ DSHARP datasets do not clearly converge on zero visibility amplitude at their longest baselines, exhibiting a mean value of $0 < {\rm Re}(V) < 1$ mJy (relative to a peak visibility amplitude of $\approx 100$ mJy). This seems to indicate that the observations are detecting a point-like source -- namely the innermost disc, whose brightness increases sharply toward $r=0$.
A \fr visibility fit strongly drives to zero once its SNR threshold is reached (which is a deliberate choice motivated by the high uncertainty in extrapolating the fit beyond the longest well-sampled baselines). And a steep slope in the fit at any baseline is represented in the brightness profile as structure on the corresponding spatial scale. Thus for a dataset that does not converge on zero at long baselines, a steep slope in the \fr fit prior to the baseline at which the visibilities converge on zero can impose false oscillations on the brightness profile. 
These oscillations manifest as a sinc-like function, at constant spatial period (the inverse of the spatial frequency location of the slope in Fourier space) and at an amplitude that diminishes away from $r=0$. 

To prevent this artifacting, we have developed an extension to \fr for a \lq{}point source-corrected model\rq{} to effectively subtract a point source from the visibilities and fit the resulting \lq{}residuals\rq{}, which are centered on Re$(V) = 0$ at long baseline. By doing this we have implicitly assumed that there is a strong point source at the center of the disc. This model is one of an infinite number of choices to extrapolate the fitted visibility distribution to inaccessible scales (a requirement of any imaging algorithm) while remaining consistent with the observed data. The choice is however sensible, as it is both physically and practically motivated. Discs are expected to rapidly increase in brightness towards the star, and applying no point-source correction can lead to spurious, coherent oscillations in the recovered brightness profile.

A pure point source (Delta function) in real space transforms to a constant visibility amplitude at all baselines. While the innermost disc is not physically a Delta function, we find this approximation works well in an unresolved component fit. In the point source-corrected model, we first subtract a constant amplitude from the visibilities, equal to the mean offset from zero at the dataset's longest baselines (specifically, those beyond the point at which the binned visibility SNR begins to oscillate about $SNR = 1$). Then we perform a standard \fr fit on the \lq{}residual\rq{} visibilities, and finally add the constant amplitude offset back into the \fr visibility fit.
Empirically, we have found this approach does a reasonable job of preventing artifacting in the \fr brightness profile for each of the $11$ DSHARP datasets whose visibilities do not clearly converge on zero (we will note these discs in Sec.~\ref{sec:methodologies}). However the technique does not fully suppress oscillations in the brightness profile in some sources, particularly in the innermost disc. In these cases the amplitude and spatial period of oscillations is sensitive to the point source amplitude; an example is shown in Sec.~\ref{sec:appendix_point_source}. We therefore assess the associated uncertainty by comparing, for each source, the fit that uses the point source amplitude as determined above with a fit that uses a $1.5\times$ larger point source amplitude (an example case is discussed in Sec.~\ref{sec:appendix_point_source}). This is motivated by a model with a larger point source amplitude effectively fitting the data to shorter baseline, which yields a more conservative estimate of small scale substructure in the brightness profile. In the main text we show the difference between the profiles of these two point source fits as an informal uncertainty band.

\subsection{Model limitations}
\label{sec:model_limitations}
The model's notable limitations in the context of this analysis are:
\begin{enumerate}
\item The 1D (axisymmetric) approach fits for the azimuthal average of the visibility data at each baseline. The model is thus inaccurate for any annulus at which the brightness is not perfectly symmetric, averaging an asymmetry over $2\pi$ in azimuth. Azimuthally localized features such as a bright arc then appear in the 1D brightness profile as a plateau or \lq{}bump\rq{} (depending on their relative brightness; we will identify specific instances). Especially for super-resolution features not seen in a \cl image, it can be difficult in some cases to distinguish the artifact of an asymmetry from an underresolved annular feature using only the 1D \fr brightness profile and observed visibilities. 

To partially resolve this ambiguity, we image the \fr fit residual visibilities to exploit that the axisymmetric model fits for the average brightness at each annulus. This effectively isolates azimuthal asymmetries in the imaged residuals\footnote{While azimuthal asymmetries are \lq{}isolated\rq{} in the imaged \fr residuals, their brightness in the image is biased because the 1D fit cannot localize flux azimuthally. The fit recovers the {\it total} flux in any annulus correctly. But a feature such as a bright arc that is localized in azimuth will have its imaged brightness biased low, because the fit distributes it over the full $2\pi$ in azimuth.}, allowing us to identify radii at which asymmetries are coincident with features in the reconstructed brightness profile. But for discs that have overlapping annular structures and azimuthal asymmetries (in DSHARP, discs with prominent spirals), interpretation is more ambiguous.
We generate a \fr residual image using the same imaging parameters as the \cl image of the source; the residual image is thus convolved and at lower resolution than the \fr brightness profile. Assessment of these residual images is therefore not a substitute for analysis with a 2D super-resolution model. 

The axisymmetric approach in \fr is also incorrect for fields of view with multiple sources (AS~205 and HT~Lup in the DSHARP sample), as these are asymmetric on large scales. Structure on the scale of a secondary disc must at some level bias the \fr fit for the primary, and we have tested the severity of this effect by refitting the HT~Lup dataset after subtracting out the secondary disc seen in the \cl image. We found this to only weakly alter the morphology of the \fr brightness profile for HT~Lup. We verified this weak sensitivity with mock datasets containing brightness asymmetries, in which we found a \fr brightness profile to be trivially altered by structure on a given scale at radii where that structure is not present. Regardless, application of the model to a field of view with multiple sources is formally incorrect.

\item While \fr produces an estimate of the uncertainty on the fitted brightness profile, the estimate is not reliable because reconstructing the brightness from Fourier data is an ill-posed problem \citep[see the discussion of this in][]{frank_method}. In particular, we do not have a robust approach for accurately extrapolating visibility amplitudes in a given dataset beyond the longest baseline that \fr fits. The uncertainty on the brightness profile produced by the model is an underestimate, and we thus do not show a formal uncertainty on any profile in this work (the uncertainty described in Sec.~\ref{sec:point_source_fits} is informal). The uncertainty on spatial scales well resolved by a \fr fit is very low as demonstrated with mock data in \citet{frank_method}. We note that the $1\sigma$ contour typically shown as an uncertainty on \cl brightness profiles is also often an underestimate, as will be evident by comparing the \cl and \fr profiles in this work. A valuable test of systematics in the extrapolation of any model is perhaps best achieved in practice by comparing observations of the same source at different resolutions (see, e.g., \citealt{Yamaguchi_2020} for this comparison using sparse modeling, or \citealt{frank_method} for such a comparison with \fr fits to moderate resolution and DSHARP observations of AS~209).  

\item The current \fr model fits for the brightness in linear space and is not positive definite (see Appendix~C in \citealt{frank_method}). Consequently the \fr brightness profile for a disc with a deep gap or an inner cavity can exhibit negative brightness in this region. We will enforce that such fits must have nonnegative brightness (which trivially affects the visibility domain fit) and will note discs for which we impose this constraint.
\end{enumerate}

\begin{figure*}
	\includegraphics[width=\textwidth]{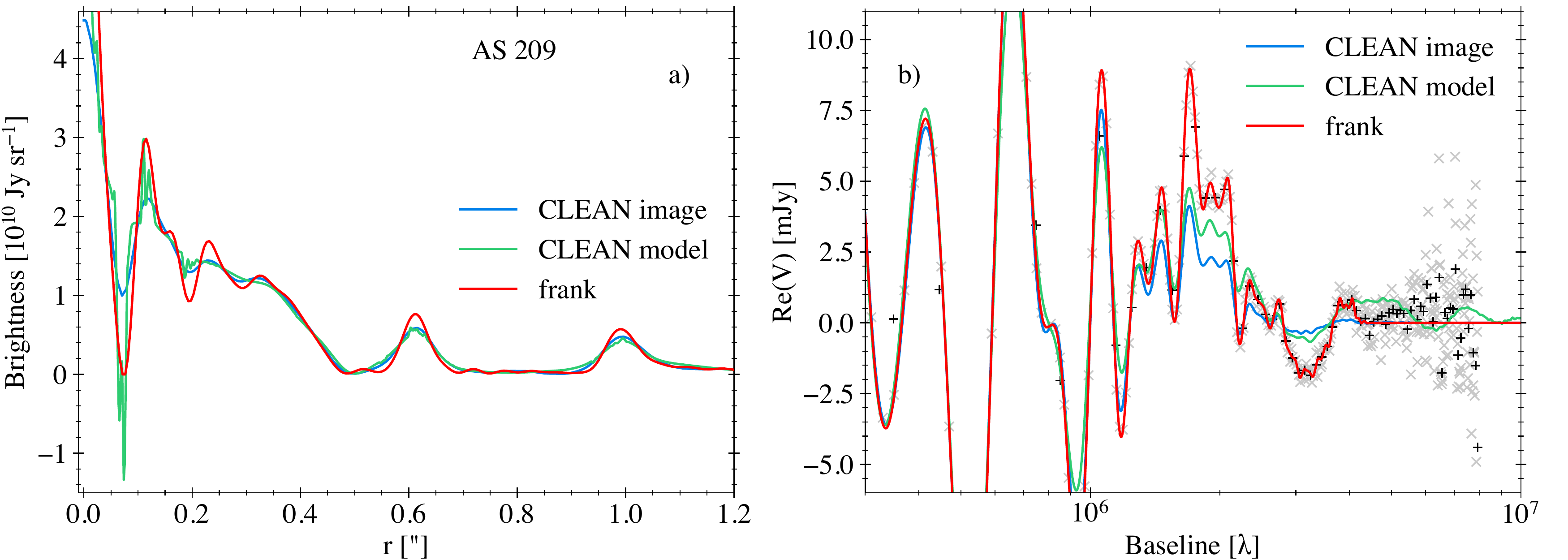}
	    \caption{{\bf Effect of \cl beam convolution (and other factors) on substructure recovery in DSHARP} \newline
	    a) For the DSHARP observations of SR~4, radial brightness profiles extracted from the \cl image and \cl model, as well as the \fr brightness profile. The \fr profile and \cl model profile peak at $8$ and $16 \times 10^{10}\ {\rm Jy\ sr}^{-1}$ respectively. \newline
	    b) The discrete Hankel transform (DHT) of the \cl profiles in (a), and the \fr visibility fit. Data are shown in $20$ and $100$ \kl bins.
	    }
    \label{fig:convolve_real_data}
\end{figure*}

\section{Methodologies -- Assessing effective \cl resolution and fit accuracy}
\label{sec:methodologies}
Here we motivate resolution limitations that affect \cl images and \cl models (Sec.~\ref{sec:convolution_theory}), then compare the accuracy of brightness profiles extracted from \cl images and \cl models to the \fr fits for all DSHARP datasets, quantifying the resolution improvement in \fr (Sec.~\ref{sec:clean_frank_accuracy}). We summarize the principles of comparing \fr fits to \cl in Sec.~\ref{sec:frank_clean_context}.

\subsection{\cl model and image resolution}
\label{sec:convolution_theory}
As noted in Sec.~\ref{sec:intro}, convolution of the \cl model image with the \cl beam induces resolution loss in the final \cl image (and thus the 1D brightness profile). As an example, convolution of a circular beam whose full-width-at-half-maximum ({\it FWHM}) is equal to the FWHM of a Gaussian feature in a brightness profile in a broadening of the feature by $\approx 40\%$ and a reduction in its amplitude by $\approx 30\%$.
Convolution in real space corresponds to multiplication in Fourier space, which induces a loss in resolution in the visibility domain via an underestimate of the observed visibility amplitudes, an effect that worsens with baseline. The FWHM of a Gaussian in real space as a function of radius $r$ corresponds to a FWHM in Fourier space as a function of spatial frequency $q$ by ${\rm FWHM_q} = 4\ln(2) /(\pi\ {\rm FWHM_r})$, obtained by relating the standard deviations in real and Fourier space.

While \cl beam convolution is the primary source of resolution loss in the \cl procedure, additional contributions can arise from, e.g., non-Gaussianity of the PSF (dirty beam). 
To assess the inherent performance of the \cl algorithm -- the resolution prior to \cl beam convolution -- it is thus useful to examine the \cl model image (the {\it .model} output of {\tt tclean}). A brightness profile extracted from this image directly measures the algorithm's achievable resolution and can itself be used to quantify a source's emission features. Some real astrophysical flux may be missed because the final residual image has not been added to the model image, and the brightness profile is often noisy due to the model image's sparse composition. But the Fourier transform of a profile extracted from the model image can quantify how well the modeling framework in the \cl procedure fits the observed visibility distribution as a function of baseline.

To this end, {\bf Fig.~\ref{fig:convolve_real_data}} compares the brightness profiles extracted from the convolved \cl image and the \cl model, as well as the Fourier transform of these profiles, for the DSHARP observations of AS~209. The profiles identify the same features in Fig.~\ref{fig:convolve_real_data}(a), but the \cl model profile shows higher amplitudes (though also more noise) and narrower widths for the two innermost disc features. This resolution advantage is not maintained across all disc features, as the \cl model profile does not recover the rings in the \cl image profile at $\approx 0.25 \arcsec$ and $0.33 \arcsec$. This is because not all of the real flux in the dirty image is incorporated into the \cl model. The \cl model profile also shows effectively identical widths and amplitudes as the \cl image profile for the two outer disc rings. Additionally and importantly, the \cl model can have negative components. 

The Fourier domain equivalents of these brightness profiles in panel (b) show how the transform of the \cl image profile underestimates visibility amplitudes with increasing severity as baseline increases, as expected from beam convolution. The transform of the \cl model critically still underestimates the visibility amplitudes between $\approx 1.6 - 3.7$~\ml, and overestimates amplitudes between $\approx 4.1 - 5.1$~\ml. This demonstrates that additional factors beyond \cl beam convolution are nontrivially limiting recovery of the full information content in the long baseline data, and thus that {\it the inherent performance of the \cl modeling framework is resolution-limited}. We emphasize that all DSHARP datasets were \texttt{CLEAN}ed by experts in the field \citep{Andrews2018, 2018ApJ...869L..42H}; these results trace practical resolution limits of \cl rather than the capability of a user. 

For reference, if we compare the observed visibilities for a given survey dataset to the Fourier transform of a brightness profile extracted from the \cl image, then convolve the data with a beam that minimizes the difference with the Fourier transform of the brightness profile, the average \cl beam width across the survey is increased by a factor of $1.16$. This simplistically treats all resolution-limiting factors in the \cl images as convolution operators, but it gives a sense of the aggregate resolution limitations in the \cl images beyond the effect of \cl beam convolution. PSF sidelobe structure and the compromise between resolution and sensitivity in the choice of the Briggs {\tt robust} parameter in {\tt tclean} are two notable resolution-limiting contributors.

For comparison to the \cl image and \cl model profiles, the \fr fit to AS~209 is also shown in Fig.~\ref{fig:convolve_real_data}. The \fr profile in panel (a) more highly resolves features seen in the \cl image profile and suggests a small bump at $\approx 0.16 \arcsec$ not present in either the \cl image profile or the \cl model profile.
In panel (b), the \fr visibility fit is correspondingly more accurate than the transforms of both the \cl profile and the \cl model beyond $\approx 1$~\ml; factors problematic for \cl such as PSF sidelobe structure are not limiting the \fr fit resolution. {\it \fr is thus outperforming the inherent resolution capability of the \cl algorithm}. This relative performance holds across the DSHARP survey, as we will now quantify.

\begin{figure*}
	\includegraphics[width=\textwidth]{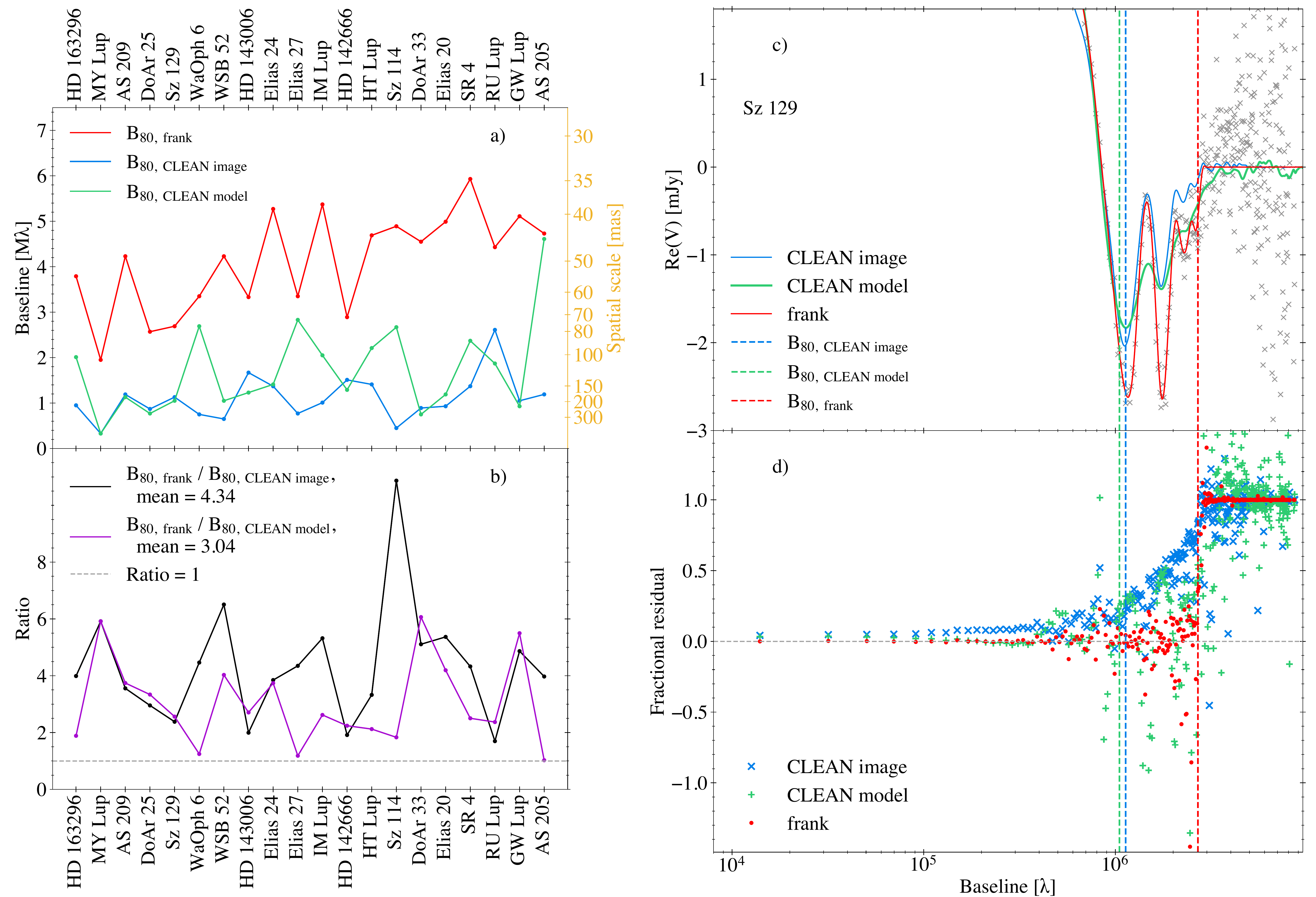}
	    \caption{{\bf \cl and \fr fit accuracies in DSHARP} \newline
	    a) Baseline accuracy metric $B_{80}$ for the convolved \cl image, \cl model, and \fr visibility fits across the $20$ DSHARP sources. The accuracy metric is the shortest baseline beyond which a fit shows $\geq 20\%$ error in visibility amplitude for a consecutive $\geq 200$ \kl (Sec.~\ref{sec:clean_frank_accuracy}). Sources are sorted by the expected baseline resolution of each dataset (see Equation~\ref{eqn:expected_resolution}). \newline
	    b) Ratio of the \fr to \cl baseline accuracy metric for both the convolved \cl image and \cl model visibility fits. \newline
        c) An example of the baseline accuracy calculation. The visibility distribution for Sz~129 ($20$ \kl bins), the \fr visibility fit, and the Fourier transform of the brightness profiles extracted from the convolved \cl image and \cl model. \newline
	    d) Fractional residuals [(data - model / data); $20$ \kl bins] for the convolved \cl image visibility fit, \cl model visibility fit and \fr visibility fit.
        }
    \label{fig:fit_accuracy}
\end{figure*}

\subsection{Using the visibilities to quantify the accuracy of \cl model, \cl image, and \fr brightness profiles}
\label{sec:clean_frank_accuracy}
It is useful to consider a metric that directly quantifies the accuracy of a 1D brightness profile extracted from a \cl image or \cl model by comparing the Fourier transform of the profile to the observed visibilities. Such a metric can incorporate all sources of error in the visibility domain representation of the profile, while being agnostic to the causes of these errors.
This metric also allows us to compare the fit accuracy in \cl and \fr. We will use as a metric a simple assessment of a profile's visibility space residuals.

We have found the most robust definition of a visibility space accuracy metric to be the shortest baseline $B_{80}$ beyond which a fit shows $\geq 20\%$ error in visibility amplitude for a minimum consecutive $200$ \kl (using $20$ \kl binning). In practice these criteria robustly identify, across all $20$ DSHARP sources, the first baseline at which the Fourier transform of a profile extracted from a \cl image or model, or the \fr visibility fit, departs appreciably from the observed visibility amplitudes and only becomes more inaccurate with increasing baseline.  
Varying the $20\%$ threshold has a weak effect on \bef, while decreasing the threshold to $10\%$ yields an average $B_{\rm 90,\ CLEAN\ image} = 0.64\ B_{\rm 80,\ CLEAN\ image}$, and $B_{\rm 90,\ CLEAN\ model} = 0.87\ B_{\rm 80,\ CLEAN\ model}$ across the $20$ DSHARP datasets. Increasing the threshold to $50\%$ gives an average $B_{\rm 50,\ CLEAN\ image} = 1.97\ B_{\rm 80,\ CLEAN\ image}$ and $B_{\rm 50,\ CLEAN\ model} = 2.26\ B_{\rm 80,\ CLEAN\ model}$.
Varying the $200$ \kl threshold has a weak effect on \bef, \bec and \bem.
The $B_{80}$ metric approximately gives a corresponding spatial scale down to which a \cl or \fr brightness profile {\it accurately} recovers substructure widths and amplitudes. A profile can of course partially recover information on smaller spatial scales, but features on these scales will be underresolved relative to the dataset's available resolution information.

{\bf Fig.~\ref{fig:fit_accuracy}}(c) -- (d) show the application of the $B_{80}$ accuracy metric to the Sz~129 DSHARP dataset.
In panel (c) the Fourier transform of a brightness profile extracted from the \cl image has some small error prior to \bec, while beyond this baseline the Fourier domain representation is, and remains, visibly inaccurate. The transform of a profile extracted from the \cl model has a \bem that is highly similar to \bec, with clear inaccuracy beyond this baseline.
Applying the same metric to determine \bef, the \fr visibility fit in Fig.~\ref{fig:fit_accuracy}(c) accurately matches the observed visibility amplitudes out to $\approx 2.8$~\ml, the baseline at which the binned data’s SNR begins to oscillate about SNR $=1$. 
Finally, the \cl (image and model) and \fr residual visibilities in Fig.~\ref{fig:fit_accuracy}(d) demonstrate the higher accuracy of the \fr fit even at moderate baselines. The \cl model residuals increase over a broad baseline range due to fundamental limitations in the \cl algorithm, while the \cl image residuals similarly increase over a broad range due additionally to \cl beam convolution. The \fr residuals remain $\approx 0$ until the sharp rise at the baseline where the fit's SNR threshold is met and the fit drives toward zero. 

\begin{table}
\caption{Expected and achieved fit accuracy metrics shown in Fig.~\ref{fig:fit_accuracy}, as well as the baseline equivalent of the data's expected resolution given in Equation~\ref{eqn:expected_resolution}}. Standard deviations assume a Gaussian distribution. Conversions to au account for the unique distance to each source. $\lambda$ is the observing wavelength; $L_{80}$ is the eightieth percentile of the baseline distribution. The last two rows give a mean and standard deviation taken across the $20$ datasets (i.e., not simply the ratio of preceding rows). 
\begin{tabular}{l c}
    \hline
    Baseline quantity, $B$ & Mean and standard deviation \\
    \hline
    $B_{\rm data,\ expected} = 0.574 \lambda / L_{80}$ & $4.75 \pm 1.39$ \ml \\
    $B_{\rm 80,\ CLEAN\ image}$ & $1.10 \pm 0.48$ \ml \\
    $B_{\rm 80,\ CLEAN\ model}$ & $1.72 \pm 0.97$ \ml \\
    $B_{\rm 80,\ frank}$ & $4.12 \pm 1.05$ \ml \\
    $B_{\rm 80,\ frank} / B_{\rm 80,\ CLEAN\ image}$ & $4.34 \pm 1.99$ \\ 
    $B_{\rm 80,\ frank} / B_{\rm 80,\ CLEAN\ model}$ & $3.04 \pm 1.47$ \\ 
    \hline
\end{tabular}
\label{tab:fit_metrics} 
\end{table}

The ordering of the baseline accuracy measurements for Sz~129 is indicative of results across the survey: $B_{\rm 80,\ CLEAN\ image} \lesssim B_{\rm 80,\ CLEAN\ model} < B_{\rm 80,\ frank}$.
Fig.~\ref{fig:fit_accuracy}(a) shows this fit accuracy analysis for all DSHARP sources, ordered by increasing \bed, the baseline equivalent of the expected angular resolution,
\begin{equation}
    \theta_{\rm data,\ expected} = 0.574 \lambda / L_{\rm 80}.
    \label{eqn:expected_resolution}
\end{equation}
Here $\lambda$ is the observing wavelength and $L_{80}$ is the eightieth percentile of the baseline distribution \citep{alma_handbook_2019}. For reference, the observed visibility distributions for the DSHARP datasets typically extend to $\approx 8 - 10$~\ml, with a mean $B_{\rm data,\ expected} = 4.72$ \ml.
Fig.~\ref{fig:fit_accuracy}(b) shows that across the $20$ DSHARP datasets, \fr is accurately fitting the visibilities to a mean factor of $4.3$ longer baseline than brightness profiles extracted from the \cl images, and a factor $3.0$ longer baseline than profiles extracted from the \cl models. This reaffirms that {\it \fr is outperforming the achieved resolution in both the \cl images and \cl models}. The resolution ratios and individual fit metrics are summarized in {\bf Table~\ref{tab:fit_metrics}}. For reference, increasing the accuracy metric's error threshold from $20\%$ to $50\%$ decreases the mean $B_{\rm 80,\ frank} / B_{\rm 80,\ CLEAN\ image}$ from $4.3$ to $3.0$, and the mean $B_{\rm 80,\ frank} / B_{\rm 80,\ CLEAN\ model}$ from $3.0$ to $1.9$.

\begin{table*}
\caption{For each DSHARP source, values for the five hyperparameters used to produce the \fr fit: SNR criterion \al, strength of smoothing \ws applied to the reconstructed power spectrum, outer radius of the fit $R_{\rm out}$, number of radial and spatial frequency points $N$ used in the fit, and floor value $p_0$ for the reconstructed power spectral mode amplitudes. Sensible choices for $R_{\rm out}$, $N$ and $p_0$ have a trivial effect on the fits: $R_{\rm out}$ is chosen to be larger than the disc's outer edge, $N$ is increased proportionally to $R_{\rm out}$, and $p_0$ is the same for all fits. Sec.~\ref{sec:model} gives a fuller explanation of, and motivation of the values for, \al and \ws. Some fits, as indicated, are forced to be nonnegative or are fit with a combined \fr and unresolved component model (in which case the visibility offset applied for the unresolved component is given); see Sec.~\ref{sec:point_source_fits}--~\ref{sec:model_limitations}. In the rightmost column, sources whose imaged \fr residuals show the brightness asymmetry discussed in Sec.~\ref{sec:geometric_effect} are noted. All \fr fits are available at \href{https://zenodo.org/record/5587841}{\color{linkcolor}{https://zenodo.org/record/5587841}}.}
\begin{tabular}{l c c c c c c c}
    \hline
    Disc & $\alpha$ & $\log_{10} w_{\rm smooth}$ & $R_{\rm out}$ [$\arcsec$] & $N$ & $p_0$ [Jy$^2$] & Fit conditions & Brightness asymmetry \\
    \hline
    AS~205 & 1.05 & -1 & 2.2 & 457 & 10$^{-15}$ & &  \\    
    AS~209 & 1.05 & -4 & 1.9 & 395 & " & Nonnegative fit & \checkmark \\  
    DoAr~25 & 1.05 & -1 & 3.1 & 500 & " & Unresolved component fit; offset $0.24$ mJy & \\
    DoAr~33 & 1.01 & -4 & 0.5 & 150 & " & Unresolved component fit; offset $0.51$ mJy & \checkmark \\
    Elias~20 & 1.01 & -4 & 1.1 & 222 & " & Unresolved component fit; offset $0.66$ mJy & \\
    Elias~24 & 1.01 & -4 & 1.9 & 395 & " & Unresolved component fit; offset $0.95$ mJy & \checkmark \\    
    Elias~27 & 1.25 & -1 & 2.9 & 500 & " & Unresolved component fit; offset $0.40$ mJy & \checkmark \\ 
    GW~Lup & 1.05 & -1 & 1.4 & 296 & " & Unresolved component fit; offset $0.73$ mJy & \checkmark \\    
    HD~142666 & 1.50 & -4 & 0.7 & 150 & " & & \checkmark \\
    HD~143006 & 1.01 & -3 & 0.8 & 173 & " & & \\
    HD~163296 & 1.01 & -4 & 2.9 & 500 & " & & \checkmark \\
    HT~Lup & 1.05 & -3 & 0.6 & 150 & " & & \\    
    IM~Lup & 1.10 & -1 & 2.4 & 494 & " & Unresolved component fit; offset $0.46$ mJy & \checkmark \\
    MY~Lup & 1.01 & -4 & 1.2 & 247 & " & Unresolved component fit; offset $0.26$ mJy & \\    
    RU~Lup & 1.05 & -4 & 0.7 & 150 & " & & \checkmark \\
    SR~4 & 1.05 & -4 & 0.5 & 150 & " & & \\    
    Sz~114 & 1.05 & -2 & 0.7 & 150 & " & Unresolved component fit; offset $0.51$ mJy & \\
    Sz~129 & 1.50 & -4 & 1.0 & 198 & " & Nonnegative fit & \\
    WaOph~6 & 1.01 & -4 & 1.9 & 395 & " & Unresolved component fit; offset $0.83$ mJy & \checkmark \\
    WSB~52 & 1.01 & -4 & 0.5 & 150 & " & Unresolved component fit; offset $0.33$ mJy & \\
    \hline
\end{tabular}
\label{tab:frank_pars} 
\end{table*}

\begin{figure*}
	\includegraphics[width=\textwidth]{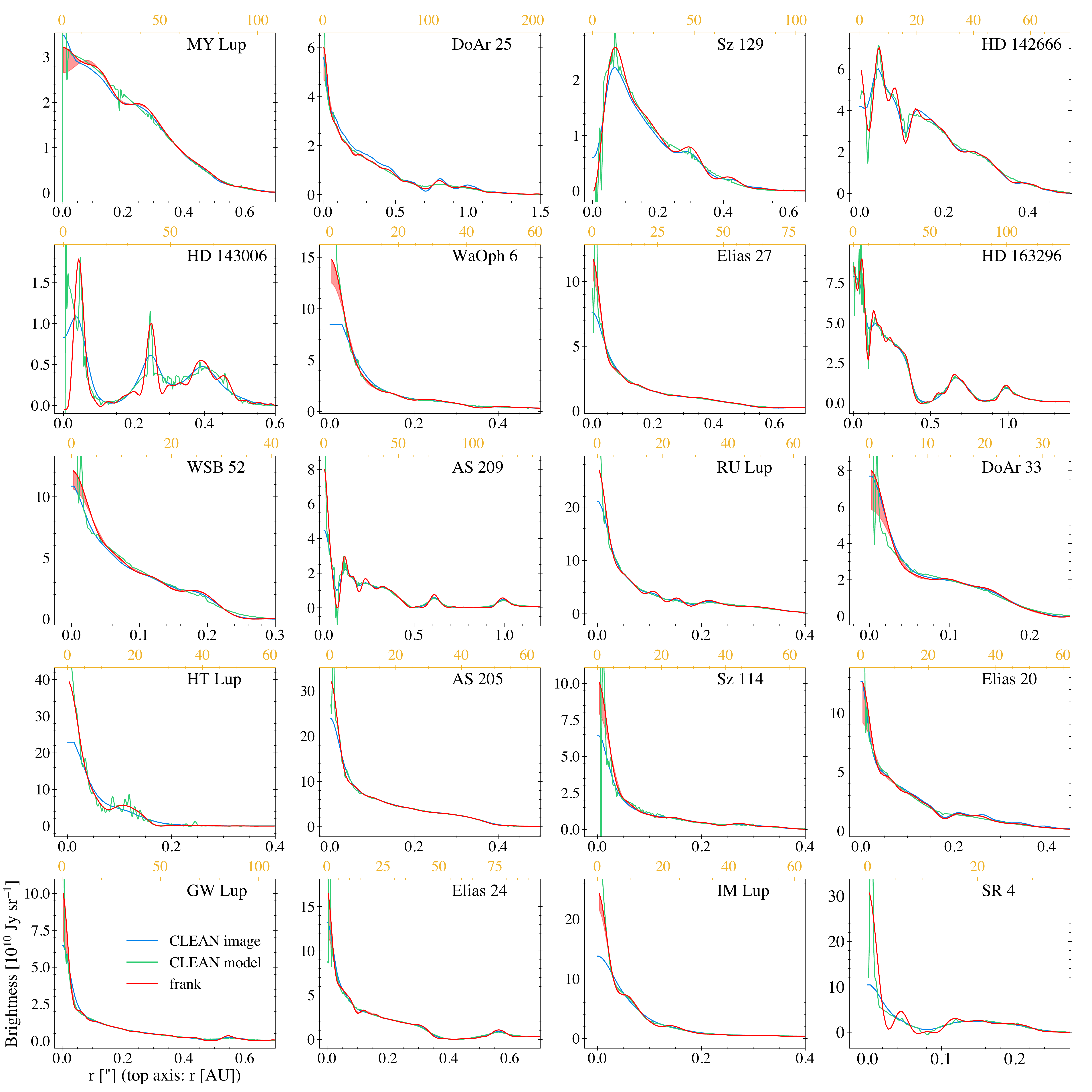}
	    \caption{{\bf Recovered brightness profiles} \newline 
	    For each source in the DSHARP survey, the convolved \cl image, \cl model and \fr brightness profiles. Some profiles zoom on the inner region of the disc. Discs are arranged from left to right and then top to bottom in ascending order of \fr fit resolution. Informal uncertainties are shown on discs fit with the point source-corrected model (Sec.~\ref{sec:point_source_fits}).
	    }
    \label{fig:profiles_grid}
\end{figure*}

\begin{figure*}
	\includegraphics[width=\textwidth]{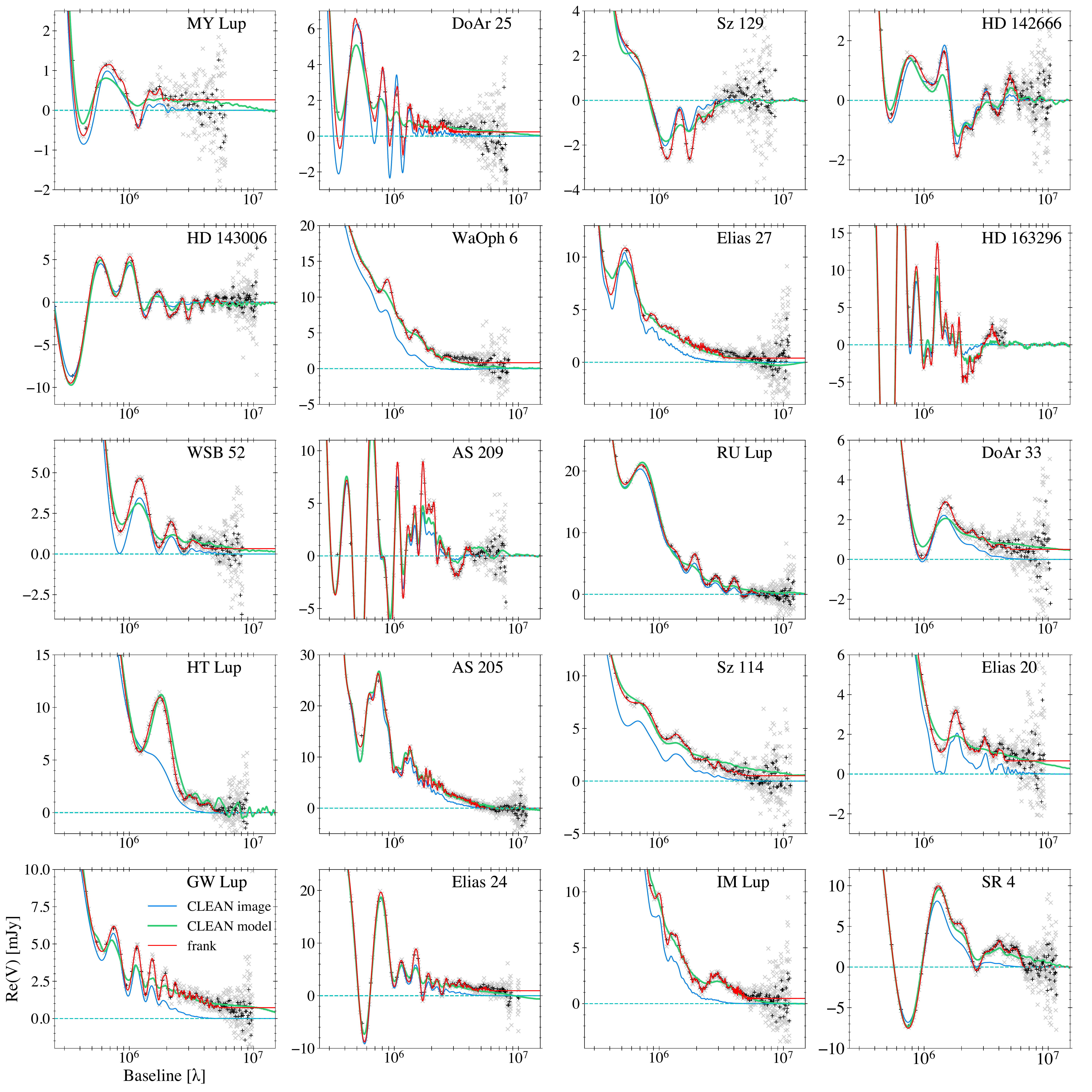}
	    \caption{{\bf Visibility fits at long baseline} \newline
	    For each source in the DSHARP survey, a zoom on the data's long baselines ($> 0.25$ \ml, corresponding to spatial scales $< 0.83 \arcsec$ mas) to show the accuracy of the \cl image, \cl model and \fr fits in matching detailed visibility structure. Data are shown in $20$ and $100$ \kl bins and become heavily noise-dominated at the longest baselines across all datasets, typically at $\gtrsim 5$ \ml. \fr does not fit these regions, as doing so would imprint noisy oscillations on the recovered brightness profile. Discs are arranged from left to right and then top to bottom in ascending order of \fr fit resolution.
        }
    \label{fig:vis_grid}
\end{figure*}

\begin{table*}
\caption{
Major new and appreciably more highly resolved features identified in \fr brightness profiles. Feature widths and gap depths are defined following the method in \citet{2018ApJ...869L..42H}; see Sec.~\ref{sec:analysis}. The datasets' finite resolution entail that the values for ring widths are upper bounds, and for gap widths and depths are lower bounds.}
\begin{tabular}{l c c c c c}
    \hline
    Disc & New (or better resolved) & Location & 
    Identifier in & Width [au] (width in & Gap depth (depth in \\
    & feature & [au] & \citet{2018ApJ...869L..42H} & \citealt{2018ApJ...869L..42H} [au]) & \citealt{2018ApJ...869L..42H}) \\
    \hline
    AS~209 & Deeper gap, & $9$ & D9 & $5.2$ ($4.7 \pm 0.2$) & $0.00$ ($0.45 \pm 0.02$) \\
    & brighter ring & $14$ & B14 & $4.9$ ($8.9 \pm 0.2$) & N/A \\  
    Elias~24 & New gap & $14$ & --- & 2.2 (---) & $0.89$ (---) \\
    GW~Lup & Deeper and sharper gap, & $75$ & D74 & $11.7$ ($12.1 \pm 0.4$) & $0.01$ ($0.31 \pm 0.03$) \\
    & brighter and narrower ring & $85$ & B85 & $7.5$ ($11.3 \pm 0.4$) & N/A \\  
    HD~142666 & New gap, & $3$ & --- & $3.6$ (---) & $0.42$ (---) \\
    & brighter ring & $7$ & B6 & $4.0$ ($5.3 \pm 1.4$) & N/A \\
    HD~143006 & Cleared inner cavity, & $\leq 7$ & --- & N/A & N/A \\
    & brighter rings, & $7$, $41$, $64$ & B6, B41, B65 & $5.3$, $5.4$, $9.5$ ($5.0 \pm 1.4$, $12.2 \pm 1.0$, $11.5 \ pm 1.4$) & N/A \\
    & wider and sharper gaps$^\dagger$, & $25, 52$ & D22, D51 & $28.4$, $16.1$ & $0.07$, $0.43$ \\ 
    & & & & ($21.7 \pm 1.0$, $12.8 \pm 1.4$) & ($0.04 \pm 0.02$, $0.53 \pm 0.02$) \\
    & brighter and narrower ring & $41$ & B41 & $5.4$ ($12.2 \pm 1.0$) & N/A \\    
    HD~163296 & Deeper gap, & $10$ & D10 & $3.0$ ($3.2 \pm 1.4$) & $0.47$ ($0.93 \pm 0.03$) \\
    & brighter ring & $13$ & B14 & $3.8$ ($3.6 \pm 1.4$) & N/A \\  
    RU~Lup & Deeper gaps & $14,\ 21,\ 29$ & D14, D21, D29 & $3.1$, $3.4$, $4.8$ & 0.90, 0.75, 0.57  \\
    & & & & (---, $<7$, $4.5 \pm 0.3$) & (---, ---, $0.78 \pm 0.01$) \\
    SR~4 & New gap, & $4$ & --- & $1.4$ (---) & $0.64$ (---) \\
    & wider and deeper gap$^\dagger$ & $11$ & D11 & $8.6$ ($6.3 \pm 1.4$) & $0.02$ ($0.23 \pm 0.02$) \\        
    Sz~129 & Cleared inner cavity, & $\leq 11$ & --- & N/A & N/A \\
    & brighter ring & $11$ & B10 & $12.3$ ($17.6 \pm 1.1$) & N/A \\    
    \hline
    \multicolumn{6}{l}{$\dagger$ Because these gaps are structured in the \fr profiles, the gap center is determined as the average of the adjacent ring centers. The gap depth is} \\
    \multicolumn{6}{l}{\ \ \ determined using the average brightness across the gap width.}
\end{tabular}
\label{tab:major_features} 
\end{table*}

\subsection{{\bf A general note on comparing \fr to \cl}}
\label{sec:frank_clean_context}
The \cl algorithm is a model to deconvolve the 2D sky brightness from the instrument PSF, which requires a functional form for the fundamental brightness unit (e.g., point sources or Gaussians). By comparison, \fr is a visibility fitter, with the express goal of accurately recovering the 1D projection of the data. This is done nonparametrically, but requires assumptions that the emission is axisymmetric and that the source geometry can be perfectly determined. These two tools can be used for different goals; in the case of accurately describing a source's azimuthally averaged brightness, \fr offers a clear resolution advantage over a profile extracted from a \cl image. The tradeoff is the potential imprint of reasonably high contrast azimuthal asymmetries on the morphology of a \fr brightness profile; this must be diagnosed by Fourier transforming (imaging) the residual \fr visibilities and/or examining the imaginary component of the observed data. 
In summary, for the purpose of obtaining a 1D brightness profile of a source (under the assumptions of axisymmetry and known source geometry), \fr will yield a more accurate (higher resolution) result, without a loss in sensitivity, compared to extracting an azimuthally averaged profile from the \cl image.
 
\section{Results}
\label{sec:result}
{\bf Fig.~\ref{fig:profiles_grid}} shows the \fr brightness profile for each DSHARP disc, as well as the \cl image profile from \citet{2018ApJ...869L..42H} and the \cl model profile obtained using the published {\tt tclean} scripts. The \fr fits exhibit more highly resolved, and in some cases new, substructure relative to the \cl images. Consistent with expectations from \cl beam convolution, the \cl image profiles also tend to underestimate the source's peak brightness (\fr must as well, albeit to a lesser extent). The \fr profiles further identify fine substructure more clearly than the noisy \cl model profiles. As a general note, feature morphologies primarily in the inner disc of the \fr profiles can be expected to evolve with higher resolution observations, which could for example find gaps to be deeper and broader, resolve rings into multiple components, or reduce the amplitude of features by placing stronger constraints on structure at the smallest scales recovered in these data. {\bf Table~\ref{tab:frank_pars}} gives the values of the hyperparameters used in each \fr fit. 

{\bf Fig.~\ref{fig:vis_grid}} shows a zoom on the long baselines of the \fr visibility fits and the Fourier transform of the \cl image and model brightness profiles across the survey.
The higher resolving power evident in the \fr brightness profiles for all $20$ sources corresponds to the \fr visibility fits matching the data at high accuracy to longer baseline than the \cl image profiles and (to a lesser extent) the \cl model profiles.
Table~\ref{tab:frank_pars} notes which \fr fits use the point source-corrected model (Sec.~\ref{sec:point_source_fits}) and gives the point source visibility amplitude applied.
For some sources -- DoAr~25, Elias~27, HD~163296, AS~205, GW~Lup, Elias~24, and IM~Lup -- \fr is clearly fitting some noise on top of the signal at long baseline. This manifests as short spatial period, low amplitude ($<1\%$ of the peak brightness) noise in the corresponding brightness profile. We accept this as a tradeoff for fitting out to baselines at which the binned data SNR approaches unity. The effect is seen most clearly in the logarithmic brightness plots for GW~Lup, Elias~24 and HD~163296 in Fig.~\ref{fig:gap_morphology} (which will be discussed in Sec.~\ref{sec:gap_morphologies}).

\section{Analysis} 
\label{sec:analysis}
{\bf Table~\ref{tab:major_features}} summarizes the major new and appreciably better resolved annular features in the \fr fits across the survey, as well as quantifies the gap/ring widths and gap depths. For the purpose of comparison, this quantification follows the approach in \citealt{2018ApJ...869L..42H} (see their \S3.2). The metric measures a gap depth as the ratio of the brightness at center of the gap $I_{\rm d}$ to the brightness at the center of the ring $I_{\rm b}$ exterior to the gap, and determines a feature width by defining the edges of an adjacent gap and ring using the average $I_{\rm mean} = 0.5 (I_{\rm d} + I_{\rm b}$). This does not yield a perfect comparison for feature widths and depths between \cl and \fr profiles, because the \fr profiles exhibit additional low amplitude substructure (e.g., in some gaps and on the wings of some rings). But as a coarse comparison, among the features in Table~\ref{tab:major_features}, $7$ of the $12$ gaps and each of the $8$ rings were quantified in \citet{2018ApJ...869L..42H}. For this subset, the \fr profiles find the gaps to be a mean $14\%$ wider and $44\%$ deeper, and the rings to be a mean $26\%$ narrower. This illustrates the utility of the super-resolution fits for substructure characterization.

Grouping the \fr brightness profiles in Fig.~\ref{fig:profiles_grid} by morphology, we can identify new substructure trends. We will exclude the multiple systems HT~Lup and AS~205 from the following analysis because, as discussed in Sec.~\ref{sec:model_limitations}, while the 1D \fr profiles are not visibly biased by the presence of multiple sources in the field of view, application of the model to such a case is still formally incorrect. We do note here that the \fr fit for HT~Lup identifies the primary disc's spiral structure as the bump in the profile at $15$ au in Fig.~\ref{fig:profiles_grid}. 

Collectively, these trends as detailed below demonstrate two broad findings. First, the DSHARP sources -- already rife with gaps and rings as identified in \citet{2018ApJ...869L..42H} -- are even more structured, especially interior to $30$ au. Second, the gaps and rings detected in the \cl images, which in many cases have widths $2 - 3 \times$ that of the \cl beam, become deeper and wider (gaps) or narrower and brighter (rings) when we fit the data with \fr.

\subsection{The compact DSHARP discs all show substructure}
\label{sec:compact_discs}
\begin{figure*}
	\includegraphics[width=\textwidth]{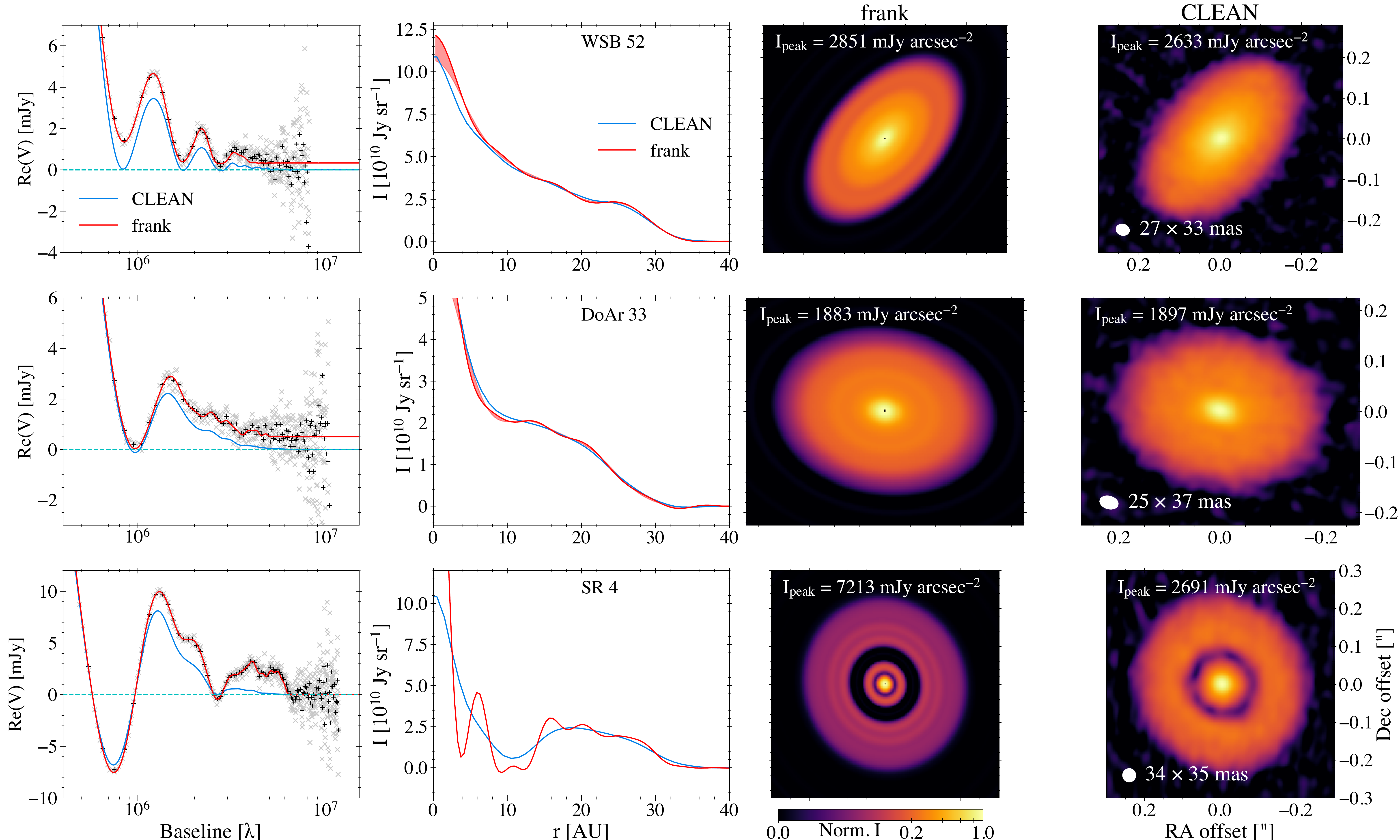}
	    \caption{{\bf Substructure in compact discs} \newline 
	    For each of the compact ($R_{\rm max} <50$ au) single-disc systems in DSHARP, a zoom on the data's long baselines ($> 0.40$ \ml, corresponding to spatial scales $< 0.52 \arcsec$ mas; data shown in $20$ and $100$ \kl bins), the \fr and \cl visibility domain fits, the \fr and \cl brightness profiles (in some cases zoomed into lower brightness), an image of the \fr profile swept over $2\pi$ and reprojected, and the \cl image. The \fr and \cl images of each disc use the same arcsinh stretch ($I_{\rm stretch} = {\rm arcsinh}(I / a)\ /\ {\rm arcsinh}(1 / a),\ a = 0.02$), but different brightness normalization. The generic color bar gives the normalized color scale, and the peak brightness is listed on each image. Discs are arranged from top to bottom by increasing \fr fit resolution. Informal uncertainties are shown on discs fit with the point source-corrected model (Sec.~\ref{sec:point_source_fits}).
        }
    \label{fig:compact_discs}
\end{figure*}

The super-resolution \fr fits find new substructure in each of the DSHARP survey's three compact ($R_{\rm max} < 50$ au), single-disc systems -- WSB~52, DoAr~33 and SR~4.
As a prominent example -- shown in {\bf Fig.~\ref{fig:compact_discs}} -- the \fr profile for SR~4 resolves the broad depression in the \cl profile into two distinct, deep gaps within $20$ au (those listed in Table~\ref{tab:major_features}). The innermost of these is centered at $4$ au; the outer, centered at $11$ au, is predicted by \fr to be at least as deep as the fit's noise floor ($\approx 10^9$ Jy sr$^{-1}$, or $4\%$ of the fitted peak brightness). 
Additionally, the \fr profile for WSB~52 finds a new, shallow gap/ring pair at $13/17$~au (in addition to the previously identified gap/ring pair at $21/25$~au), and the \fr fit for DoAr~33 resolves the single gap/ring pair at $9/17$~au in the \cl profile into two gap/ring pairs.

Typical of current observations of compact discs, the shallow features in the \fr profiles for these compact sources could be either intrinsically wide and shallow or narrow and underresolved.
Sensitive observations at higher angular resolution are needed to distinguish between the two scenarios.
We use a point source-corrected fit for WSB~52 and DoAr~33 (Sec.~\ref{sec:point_source_fits}), with the profile's sensitivity to the point source visibility amplitude shown as the informal uncertainty band in Fig,.~\ref{fig:compact_discs}. The substructure in both sources is robust to this informal uncertainty.

The commonality of substructure \fr finds across these three compact DSHARP sources suggests that in general compact discs, just as more extended discs, may routinely exhibit annular substructure. SR~4 is particularly notable in this context, with its effectively empty gap at $11$ au analogous to the empty gap \fr finds at $10$ au in the much larger disc of AS~209 (outer radius $\approx 150$ au). 
If compact discs are frequently structured, it may follow that the same physical processes (including companions) responsible for structure in larger discs are also efficacious in smaller discs. 
The improved identification of substructure in the compact DSHARP discs is also of particular interest, as compact sources represent a significant yet understudied component of the protoplanetary disc population.

\subsection{Extended discs show brighter rings, deeper gaps, and hints of inner disc substructure}
\label{sec:extended_discs}
\begin{figure*}
	\includegraphics[width=\textwidth]{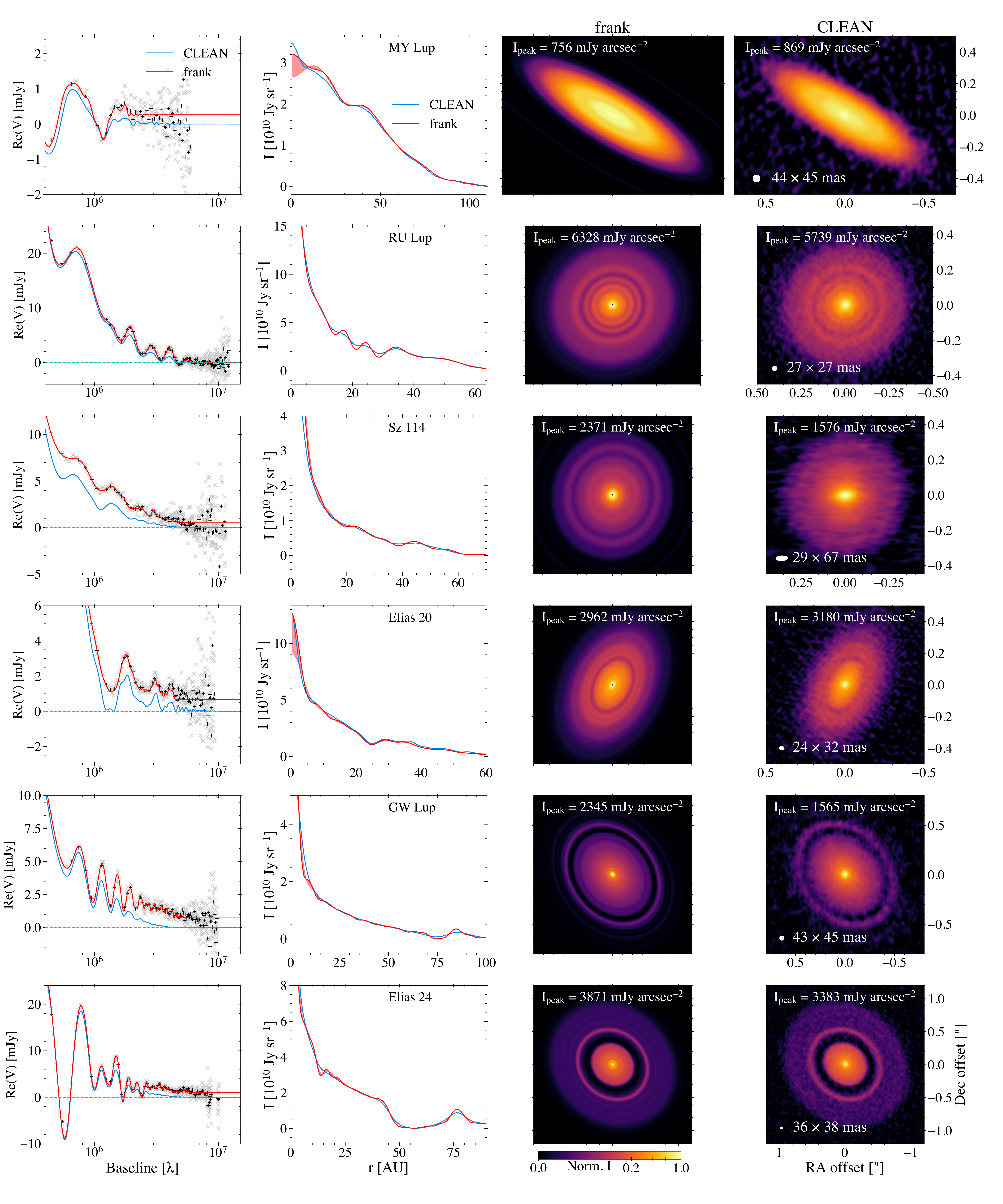}
	    \caption{{\bf Substructure in extended discs} \newline 
	    \ As in Fig.~\ref{fig:compact_discs}, but for the extended ($>50$ au) DSHARP discs in Sec.~\ref{sec:extended_discs}.
        }
    \label{fig:extended_discs}
\end{figure*}

\fr fits for several extended DSHARP sources better resolve the gaps and rings that appear shallow in the \cl profiles, as shown in {\bf Fig.~\ref{fig:extended_discs}}. This is especially apparent in the outer gap and ring pair in GW~Lup, where in the \fr profile the brightness contrast between the gap and ring is $0.01$, compared to $0.31$ in the \cl profile (see Table~\ref{tab:major_features}); and in RU~Lup, where the three consecutive gaps interior to $30$ au are deeper in the \fr profile (the contrast of the gap at $29$~au for example is $0.57$ in the \fr fit, compared to $0.78$ in the \cl profile). The \fr fit to Elias~24 robustly finds a new gap at $13$ au, and the model better resolves the faint ring at $45$ au in Sz~114.

For RU~Lup, Sz~114, Elias~20, GW~Lup, and Elias~24, the model suggests a steep inner disc in the inner $5 - 7$ au, followed by a shallower slope at slightly larger radii. This may be an indication of underresolved substructure between $\approx 7 - 12$ au. We use the point source-corrected fit (Sec.~\ref{sec:point_source_fits}) for $5$ of the $6$ sources in Fig.~\ref{fig:extended_discs} and show the profile's sensitivity to the point source visibility amplitude as the informal uncertainty band. This suggests we should be cautious about the fit's exact structure in the innermost disc, while the change in slope is robust to this uncertainty.

In addition to these sources, the \fr brightness profile for a majority of the $20$ DSHARP discs exhibits either gap and ring substructure interior to $30$ au, or clear change in slope interior to $\approx 12$ au. This suggests substructure is common not only at $\geq 30$ au, but also at the smaller separations that harbor the bulk of the observed exoplanet population. The Gaussian kernel density estimate for gap and ring locations in \citet{2018ApJ...869L..42H} peaks at $30$ au, while by comparison the \fr fits suggest that the occurrence rate continues to rise toward $r = 0$.
The (effectively) empty gaps at $\approx 10$ au in the \fr fits for AS~209 (gap contrast of $0.00$ in the \fr profile, compared to $0.45$ in the \cl profile) and SR~4 (contrast of $0.02$ in the \fr profile, compared to $0.23$ in the \cl profile) suggest that the lack of such deep features identified thus far in high resolution disc observations is an artifact of resolving power, rather than an intrinsic absence of cleared gaps in inner discs. 

\subsection{Two of the oldest DSHARP discs appear to have inner cavities}
\label{sec:inner_cavities}
\begin{figure*}
	\includegraphics[width=\textwidth]{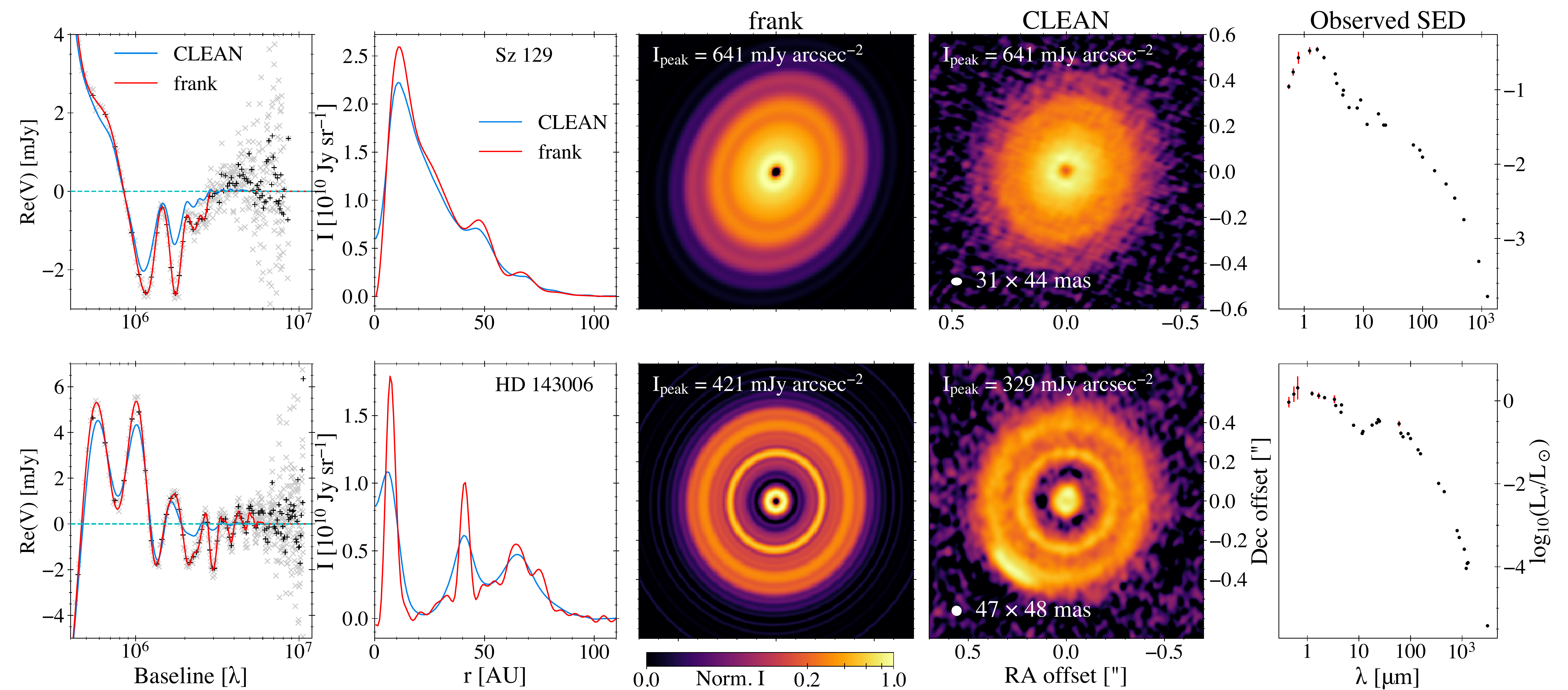}
	    \caption{{\bf Evidence for inner cavities} \newline 
	    As in Fig.~\ref{fig:compact_discs}, but for the DSHARP discs showing indications of inner cavities. Additionally shown are the observed spectral energy distributions \citep{Andrews2018}. The azimuthally localized bright arc along the outer edge of the outer ring in the \cl image for HD~143006 is erroneously visualized as a symmetric feature in the \fr image (because the model is 1D) and manifests in the \fr brightness profiles as the \lq{}bump\rq{} at $77$ au.
        }
    \label{fig:inner_cavities}
\end{figure*}

\fr finds that $2$ of the $20$ DSHARP discs, HD~143006 and Sz~129, have a fully cleared inner cavity. The \cl profiles for these sources show a decreasing brightness toward $r=0$, but not a full cavity in {\bf Fig.~\ref{fig:inner_cavities}}, and the \fr fits also find the discs to have an appreciably brighter inner rim (noted in Table~\ref{tab:major_features}). \citet{2018ApJ...869L..42H} inferred the presence of a cleared cavity in these sources from the \cl images, now confirmed by the \fr fits.
The spectral energy distribution (SED) for HD~143006 (and potentially for Sz~129) shows a dearth in the near-IR ($\approx 10 - 20\ \mu{\rm m}$) and excess in the far-IR ($\approx $20$ - 100\ \mu{\rm m}$) as shown in Fig.~\ref{fig:inner_cavities} \citep[SEDs adapted from][]{Andrews2018}. These may be indications of transition discs; it is also possible that either of these sources has a sharp rise in brightness in the innermost disc that is not resolved by \fr.

Intriguingly, HD~143006 and Sz~129 may be two of the oldest discs in the DSHARP sample. Among the survey's single-disc systems, $5$ of $18$ orbit a star whose inferred age is $>2$ Myr as reported in \citealt{Andrews2018} (see specific references in their Table~1): HD~143006 ($4.0 \pm 2.0$ Myr), Sz~129 ($4.0 \pm 2.5$ Myr), MY~Lup ($10.0^{+4.0}_{-2.0}$ Myr), HD~142666 ($12.6 \pm 0.3$ Myr), and HD~163296 ($12.6 \pm 4.0$ Myr). These estimates are in general subject to systematic challenges such as interpreting robust ages at high effective temperature, and \citet{Andrews2018} additionally note that the age for MY~Lup may be overestimated due to the inclined and flared disc extincting the stellar spectrum. Of the remaining four potentially old sources, HD~143006 and Sz~129 show inner cavities in the \fr fits, while HD~142666 and HD~163296 both show gaps interior to $\approx 5$ au. No other \fr brightness profile in DSHARP shows a turnover in brightness interior to $5$ au, which may tentatively suggest that these four objects are experiencing the later stages of disc dispersal, losing (or having already lost) their inner disc at their potentially advanced ages. The expectation is in line with the finding by \citet{2014prpl.conf..497E} that the fraction of transition discs in star forming regions and young clusters increases from $\approx 1\%$ to $\approx 10\%$ for ages $\gtrsim 2$ Myr (these percentages do carry large uncertainties). 

More speculatively, HD~142666, HD~143006 and HD~163296 are $3$ of the $4$ most structured discs in the survey, perhaps indicating that even if annular substructures do form early, discs may become more structured over time (e.g., as additional planets form). AS~209 complicates this interpretation though, being the other highly structured disc in the survey and having an inferred age of only $1.0^{+2.5}_{-1.0}$ Myr. 

\subsubsection{Improved constraints on dust trapping}
\label{sec:dust_trap}
The narrower rings in the \fr fits relative to \cl can offer improved constraints on dust trapping. \citet{2018ApJ...869L..46D} examine the outer disc rings in the \cl profiles for five of the DSHARP sources -- AS~209, Elias~24, GW~Lup, HD~143006, and HD~163296 -- and infer deconvolved widths $w_{\rm dust}$ to compare to the local pressure scale height $h_{\rm p}$. If this ratio is $<1$, the rings are inferred to be the result of dust traps. With this ratio a plausible range of widths for gas pressure bumps $w_{\rm gas}$ at the radial location of the dust rings can also be determined, in turn yielding a range of values for the ratio of the viscosity parameter to the local Stokes number (\citealt{2018ApJ...869L..46D}, Equation~21), 
\begin{equation}
 \frac{\alpha_{\rm turb}}{{\rm St}} = \Bigg[\Big(\frac{w_{\rm gas}}{w_{\rm dust}}\Big)^2 - 1\Bigg]^{-1}.
\end{equation}
The lower this ratio, the lower the threshold to induce the streaming instability.
\citealt{2020MNRAS.495..173R} take a similar approach, using the dust ring widths together with deviations from Keplerian velocity inferred from the $^{12}$CO observations in AS~209 and HD~163296 to measure $\alpha_{\rm turb} / {\rm St}$. According to their Equation~1,
\begin{equation}
 \frac{\alpha_{\rm turb}}{{\rm St}} = -\frac{2w_{\rm dust}^2}{r_0} \frac{v_{\rm k}^2}{c_{\rm s}^2} \frac{{\rm d}}{{\rm dr}} \Big(\frac{\delta v_\phi}{v_{\rm k}}\Big).
\end{equation}
Here $r_0$ is the radial location of the dust ring, $v_{\rm k}$ the local Keplerian velocity, $c_{\rm s}$ the sound speed, and $\delta v_\phi = v_\phi - v_{\rm K}$ is the deviation from Keplerian.

Following the procedure in \citet{2018ApJ...869L..46D} to determine dust ring widths, we find each of the $8$ rings in the \fr profiles are narrower than even the deconvolved widths in \citet{2018ApJ...869L..46D}, by a mean $24\%$. The \fr widths are also narrower than the $4$ of these rings examined in \citet{2020MNRAS.495..173R} by a mean $13\%$. 
{\bf Table~\ref{tab:dust_trap}} compares the \fr widths to those in \citet{2018ApJ...869L..46D} and \citet{2020MNRAS.495..173R}, as well as the corresponding estimates of $w_{\rm dust} / h_{\rm p}$.
The narrower \fr dust rings yield a reduction in estimates of $\alpha_{\rm turb} / {\rm St}$ by a mean $47\%$ relative to \citealt{2018ApJ...869L..46D} (for $w_{\rm min}$, the minimum width of the gas pressure bump; see that work) and by a mean $25\%$ relative to \citet{2020MNRAS.495..173R}.
These results suggest the dust ring widths in \citet{2018ApJ...869L..46D} and \citet{2020MNRAS.495..173R} are overestimates, and that smaller values of $\alpha_{\rm turb}$ (or larger values of ${\rm St}$) are thus needed to agree with the true (unknown) ring widths. A smaller ratio of $\alpha_{\rm turb} / {\rm St}$ would in turn correspond to a lower threshold for inducing the streaming instability.

To emphasize the importance of an accurate visibility fit, we note that \citet{2018ApJ...869L..46D} find the deconvolved ring widths are in some cases wider, but in others narrower, than the widths determined by parametrically fitting the visibilities for AS~209 \citep{2018ApJ...869L..48G}, HD~163296 \citep{2018ApJ...869L..49I} and HD~143006 (\citealt{2018ApJ...869L..50P}; see Appendix~C in \citealt{2018ApJ...869L..46D}). 
The \fr profiles instead yield narrower rings than the deconvolved widths in \citealt{2018ApJ...869L..46D} in all cases, because \fr is fitting structure in the observed visibilities to longer baseline than the parametric visibility fits.
Comparing the \fr visibility fit for HD~163296 to the parametric visibility fit in \citealt{2018ApJ...869L..49I} for example, 
\fr accurately traces the visibilities to $\approx 3.8$ \ml, while the parametric fit begins to show clear error beyond $\approx 0.9$ \ml, and the \fr ring widths are thus narrower.  

\begin{table*}
\caption{Dust trapping constraints from \fr rings (see Sec.~\ref{sec:dust_trap}). 
Column (1): Disc name. 
(2): Ring name in \citet{2018ApJ...869L..42H}. 
(3): Measured \fr dust ring widths $w_{\rm dust,\ frank}$, deconvolved widths $w_{\rm dust,\ decon.}$ \citep{2018ApJ...869L..46D}, and widths inferred using the $^{12}$CO rotation curve $w_{\rm dust,\ rot.\ curve}$ \citep{2020MNRAS.495..173R}. 
(4): Ratio of the ring widths in (3) to the  pressure scale height $h_{\rm p}$. (5): Ratio of turbulent viscosity to Stokes number $\alpha_{\rm turb} / {\rm St}$, using minimum gas pressure bump widths $w_{\rm gas,\ min.}$ following \citet{2018ApJ...869L..46D}. For cases in which $w_{\rm dust} / h_{\rm p} \geq 1$, values of $\alpha_{\rm turb} / {\rm St}(w_{\rm gas,\ min.})$ are not given.
(6): Ratio of turbulent viscosity to Stokes number $\alpha_{\rm turb} / {\rm St}$, using gas pressure bump widths $w_{\rm gas,\ rot.\ curve}$ following \citet{2020MNRAS.495..173R}.
Widths $w$ in columns (3) -- (6) are defined as the standard deviation of a Gaussian. 
}
\begin{tabular}{l c c c c c}
    \hline
    Disc & Ring & $w_{\rm dust,\  frank}$ [au] & $w_{\rm dust,\ frank} / h_{\rm p}$ & $\alpha_{\rm turb} / {\rm St}(w_{\rm gas,\ min.,\ frank})$ & $\alpha_{\rm turb} / {\rm St}(w_{\rm gas,\ rot.\ curve,\ frank})$ \\
    & identifier & ($w_{\rm dust,\ decon.}$ [au]) & ($w_{\rm dust,\ decon.} / h_{\rm p}$) & $\big(\alpha_{\rm turb} / {\rm St}(w_{\rm gas,\ min.,\ decon.})\big)$ & \{$\alpha_{\rm turb} / {\rm St}(w_{\rm gas,\ rot.\ curve})$\} \\
    & & \{$w_{\rm dust,\ rot.\ curve}$ [au]\} & \{$w_{\rm dust,\ rot.\ curve} / h_{\rm p}$\} & \\ 
    (1) & (2) & (3) & (4) & (5) & (6) \\
    \hline
    AS~209 & B74 & $2.86$ ($3.38$) \{$3.39 \pm 0.06$\} & $0.5$ ($0.6$) \{$0.6$\} & $0.35$ ($0.57$) & $0.13$ \{$0.18 \pm 0.04$\}\\
    AS~209 & B120 & $3.63$ ($4.11$) \{$4.12 \pm 0.07$\} & $0.4$ ($0.4$) \{$0.4$\} & $0.14$ ($0.19$) & $0.10$ \{$0.13 \pm 0.02$\}\\
    Elias~24 & B77 & $3.41$ ($4.57$) & $0.5$ ($0.6$) & $0.29$ ($0.66$) & \\
    HD~163296 & B67 & $6.32$ ($6.84$) \{$6.85 \pm 0.03$\} & $1.5$ ($1.6$) \{$1.6$\} & --- (---) & $0.19$ \{$0.23 \pm 0.03$\} \\
    HD~163296 & B100 & $3.80$ ($4.67$) \{$4.66 \pm 0.08$\} & $0.5$ ($0.7$) \{$0.7$\} & $0.40$ ($0.77$) & $0.03$ \{$0.04 \pm 0.01$\} \\
    GW~Lup & B85 & $3.12$ ($4.80$) & $0.4$ ($0.6$) & $0.21$ ($0.68$) & \\
    HD~143006 & B41 & $2.09$ ($3.90$) & $1.0$ ($1.9$) & --- (---) & \\
    HD~143006 & B65 & $4.99$ ($7.31$) & $1.4$ ($2.0$) & --- (---) & \\
    \hline
\end{tabular}
\label{tab:dust_trap} 
\end{table*}

\subsection{Spiral arms appear to extend into the spiral discs' cores}
\label{sec:spiral_discs}
\begin{figure*}
	\includegraphics[width=\textwidth]{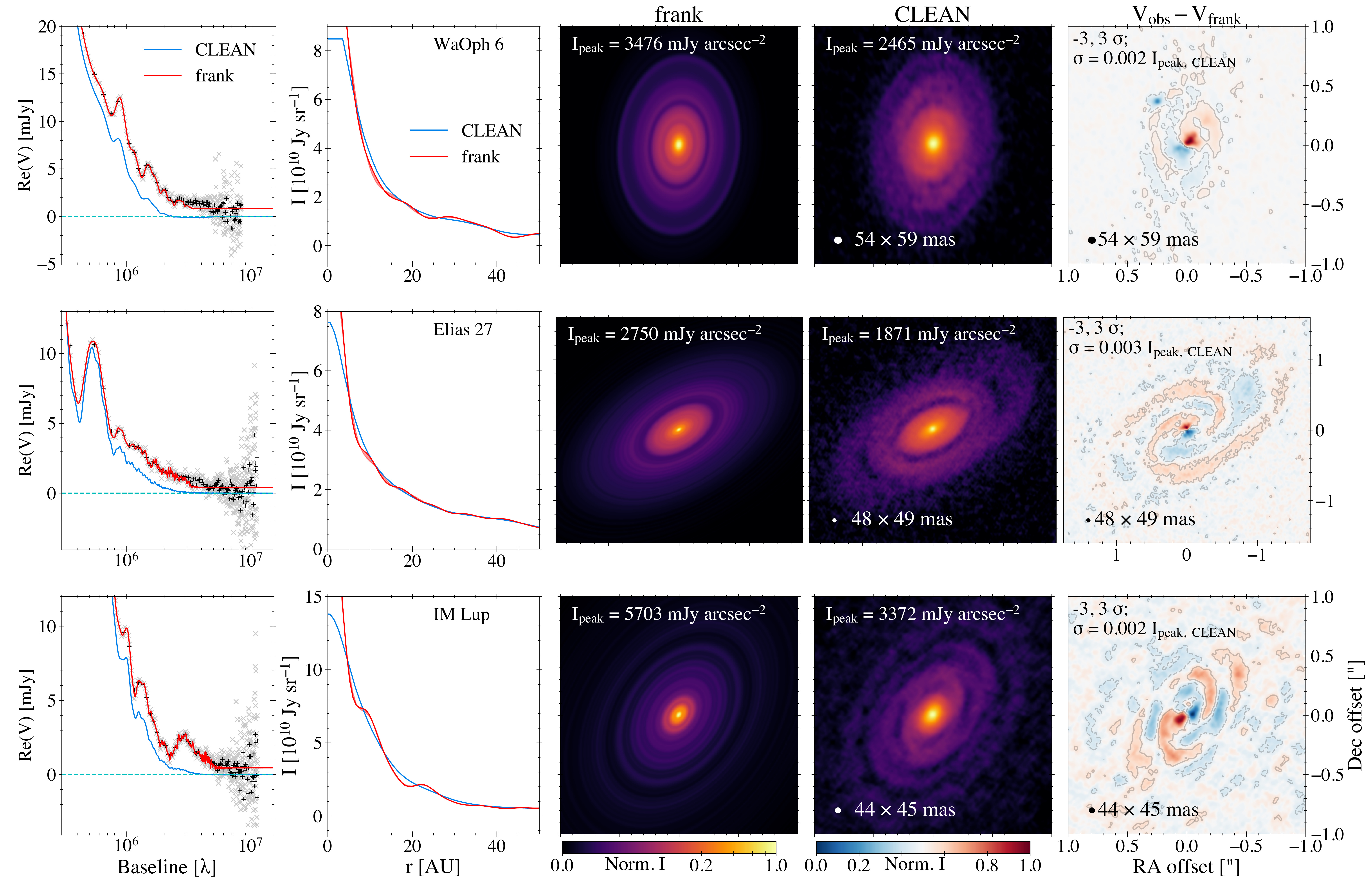}
	    \caption{{\bf Tracing spiral arms into their disc's cores} \newline 
	    As in Fig.~\ref{fig:compact_discs}, but for the DSHARP discs exhibiting strong spiral structure. 
	    The visibility plots here zoom on baselines $> 0.30$ \ml (corresponding to spatial scales $< 0.69 \arcsec$). 
	    Additionally shown are the \fr residual visibilities imaged (0 \cl iterations). Residual images use a linear color scale (a normalized color bar is shown, and the $\sigma$ value for each image is given). Azimuthal asymmetries in \cl images are erroneously visualized as symmetric features in the \fr images because the \fr model is 1D.
        }
    \label{fig:spirals}
\end{figure*}

The \fr fits to the three single-disc systems in the survey exhibiting prominent spirals -- WaOph~6, Elias~27 and IM~Lup -- show clear deviations from a smooth envelope in the discs' bright cores, which extend to $\approx 45, 60$ and $30$ au respectively.
The imaged \fr residual visibilities\footnote{As discussed in Sec.~\ref{sec:model_limitations}, an azimuthally averaged \fr brightness profile is erroneous for any radius at which the brightness is not symmetric. However because \fr correctly fits for the {\it averaged} brightness in each annulus, subtracting the fit from the observed visibilities effectively isolates asymmetric structure in a residual image \citep[analogous to the same procedure with \cl fits in Figure 1 of][]{2018ApJ...869L..43H}.} in {\bf Fig.~\ref{fig:spirals}} suggest these features may not be tracing symmetric gaps and rings, but instead the (azimuthally averaged) innermost components of the spiral arms. This interpretation is tentatively supported by examining polar projections of the deprojected \fr imaged residuals (not shown), which appear to faintly trace the arms to moderately smaller radii than the polar plots in \citet{2018ApJ...869L..43H}.  

The model for each of these discs uses the point source-corrected fit (Sec.~\ref{sec:point_source_fits}), with the profile's sensitivity to the point source visibility amplitude shown as the informal uncertainty band in Fig,.~\ref{fig:spirals}. The exact structure in the discs' cores should thus be taken with caution, though the features in WaOph~6 beyond $\approx 20$ au, in Elias~27 beyond $\approx 15$ au, and throughout the inner disc in IM~Lup are robust to this informal uncertainty.

\subsection{The most structured DSHARP sources have morphologically similar inner discs}
\label{sec:high_structure_discs}
\begin{figure*}
	\includegraphics[width=\textwidth]{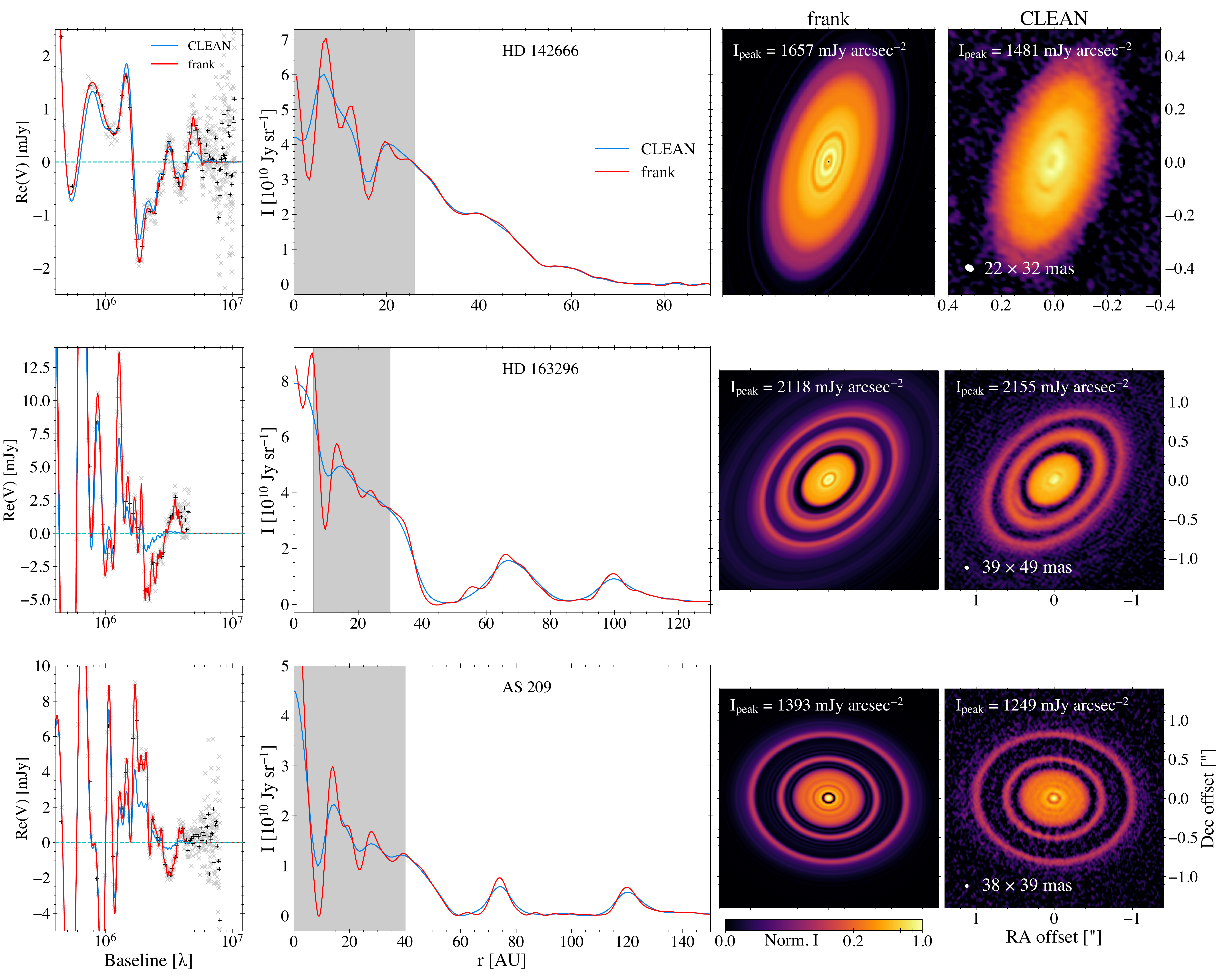}
	    \caption{{\bf Highly structured discs} \newline 
	    As in Fig.~\ref{fig:compact_discs}, but for the DSHARP discs exhibiting the highest density of substructures. The azimuthally localized bright arc along the inner edge of the intermediate ring in the \cl image for HD~163296 is erroneously visualized as a symmetric feature in the \fr image (because the model is 1D) and manifests in the \fr brightness profile as the \lq{}bump\rq{} at $55$ au. The shaded regions show morphological similarities across discs as discussed in Sec.~\ref{sec:high_structure_discs}.
        }
    \label{fig:high_structure_discs}
\end{figure*}

\fr fits to the three most highly structured DSHARP discs --  HD~163296, AS~209 and HD~142666 -- in {\bf Fig.~\ref{fig:high_structure_discs}} more fully resolve gaps and rings present in the \cl profiles, especially the gap-ring pair in each source interior to $15$ au (noted in Table~\ref{tab:major_features}). The \fr profiles also show new substructure in the inner disc of each source that is strikingly similar: a gap-ring pair, immediately exterior to which is a gap that shows a brightness excess (potentially a pressure bump) on both of its wings, and exterior to this a shallow depression (this region is highlighted for each source in Fig.~\ref{fig:high_structure_discs}). Whether this morphological similarity, including the newly identified features, is due to the same physical process, e.g., an embedded planet, would require detailed hydrodynamic simulations that are beyond the scope of this work.

\subsection{Deep gap morphologies in \fr profiles potentially indicate embedded planets}
\label{sec:gap_morphologies}
\begin{figure*}
	\includegraphics[width=\textwidth]{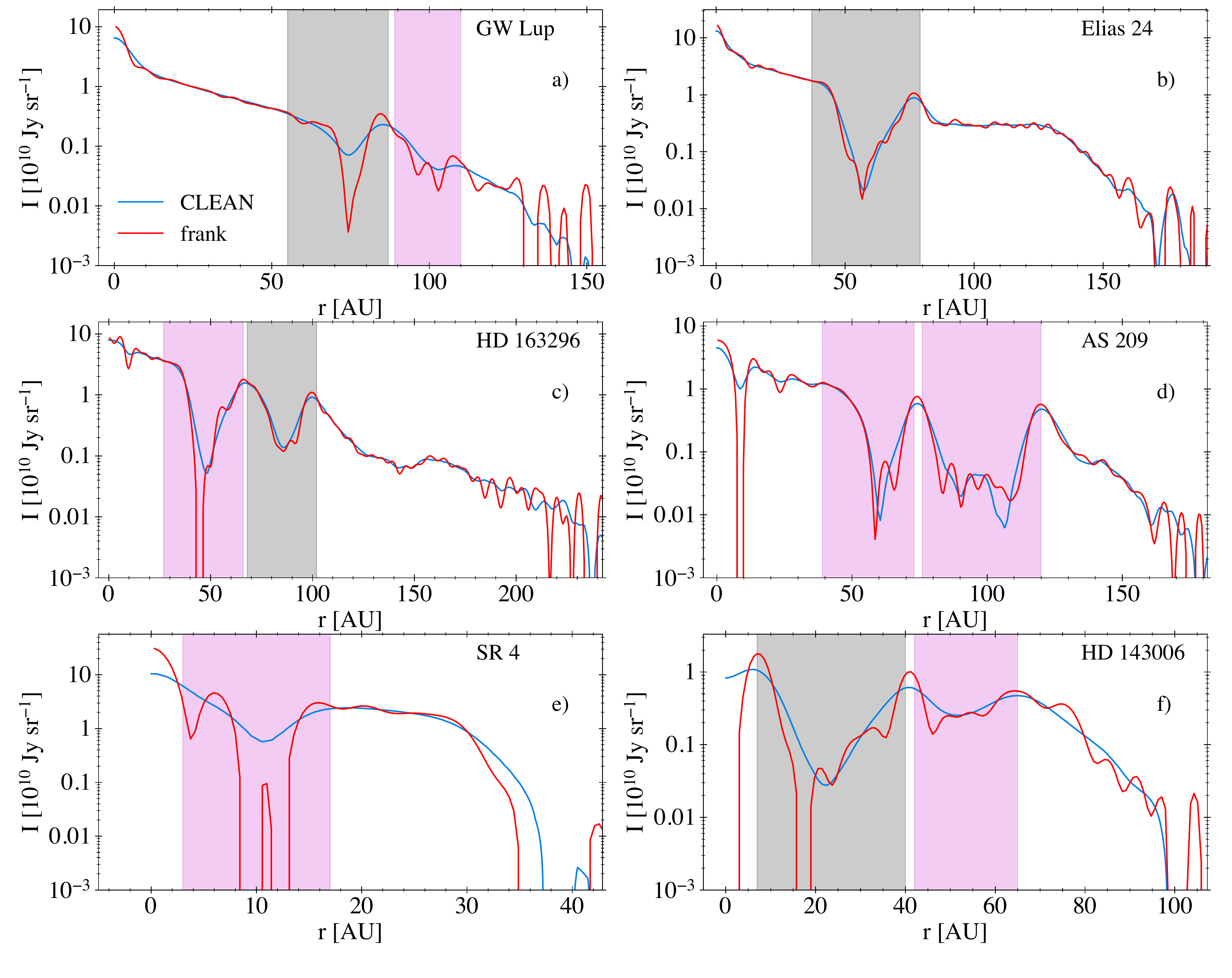}
	    \caption{{\bf Morphologies for deep and structured gaps} \newline
         \fr and \cl brightness profiles in logarithmic brightness for DSHARP discs whose \fr profiles have gaps that are either appreciably deeper or contain more structure than seen in the \cl profiles.
         Gap regions are shaded for identification; those shaded in gray have a claimed planetary detection (either from gas kinematics or direct imaging), and those in pink have no detection.
        }
    \label{fig:gap_morphology}
\end{figure*}

The \fr brightness profiles for the six DSHARP discs shown in {\bf Fig.~\ref{fig:gap_morphology}} -- GW~Lup, Elias~24, HD~163296, AS~209, SR~4, and HD~143006 -- show that deep gaps which were already prominent in the \cl profiles become deeper and/or wider with sharper edges, as well as more structured in some cases. The detailed structure within the gaps in the \fr profiles varies weakly as the fit's SNR criterion is varied (recall that we have accepted some low amplitude, short spatial period noise in the profiles as a tradeoff for fitting the visibilities out to baselines at which the binned data SNR approaches unity). Insensitive to the exact fit is the presence of local maxima exterior to the gaps, as well as less prominent maxima or shallow slopes interior to the gaps. Some of the gap morphologies (both the structure within the gap and on its edges) are qualitatively similar to the dust surface density distribution surrounding a gap-opening planet in hydrodynamic simulations (particularly those for a stationary or slowly migrating planet in \citealt{10.1093/mnras/sty2847} and \citealt{2019MNRAS.485.5914N}). However detailed simulations would be required to confirm agreement in any individual case; we leave this to a future work. The four gaps shaded in gray in Fig.~\ref{fig:gap_morphology} have a claimed planet detection: in GW~Lup \citep{2020ApJ...890L...9P}, Elias~24 \citep{2020arXiv201210464J} and both gaps in HD~163296 \citep{2018ApJ...860L..12T, Pinte_2018}; the gaps shaded in pink do not have a detection.

\subsection{A geometric viewing effect traces disc vertical structure}
\label{sec:geometric_effect}
\begin{figure*}
	\includegraphics[width=\textwidth]{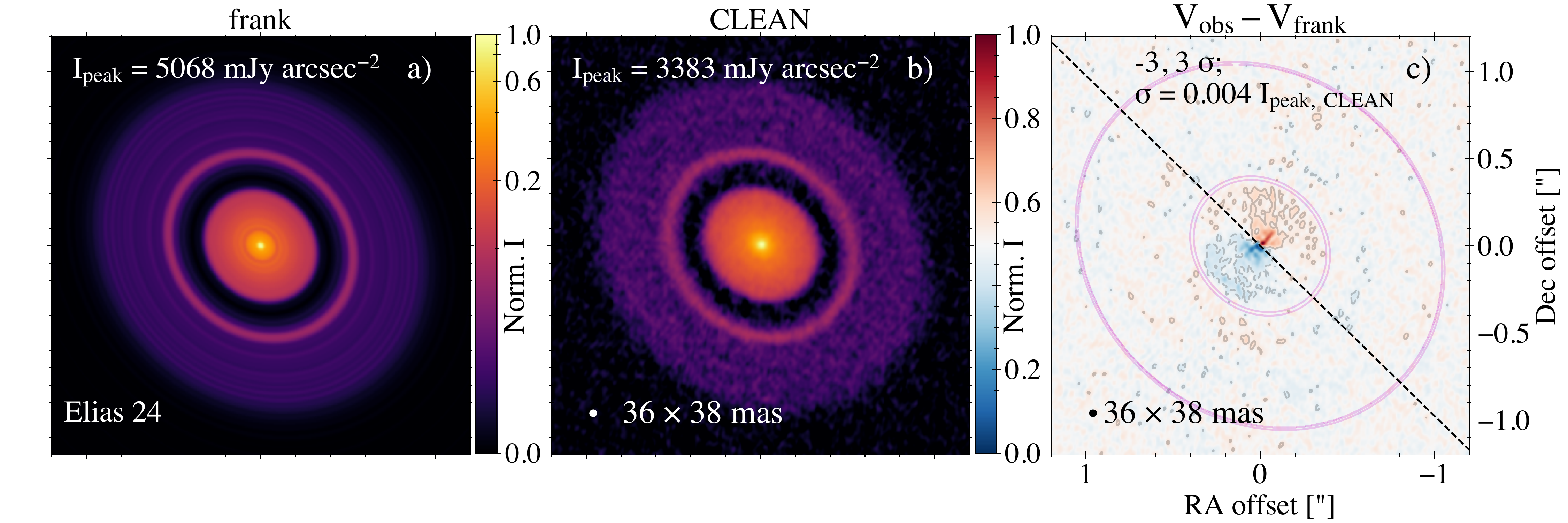}
	    \caption{{\bf A geometric viewing effect tracing disc vertical structure} \newline 
	    a) For Elias~24, an image of the \fr profile swept over $2\pi$ and reprojected. \newline
	    b) The \cl image. The \fr and \cl images of each disc use the same arcsinh stretch ($I_{\rm stretch} = {\rm arcsinh}(I / a)\ /\ {\rm arcsinh}(1 / a),\ a = 0.02$), but different brightness normalization. The generic color bar gives the normalized color scale, and the peak brightness is listed on both images. \newline 
	    c) The \fr residual visibilities imaged (0 \cl iterations), with contours overplotted, as well as additional lines tracing the outer edge of the inner disc and the disc outer edge (from (a)), and a dashed line along the fitted position angle (as a proxy for the disc's major axis). The residual image is convolved with the published \cl beam and uses a linear color scale. The shown $3 \sigma$ contours correspond to a residual brightness $<1\%$ of the local average brightness in the \cl image at the outer edge of the inner disc, $42$ au. The residual image uses a linear color scale (a normalized color bar is shown, and the $\sigma$ value for each image is given).
        }
    \label{fig:geometric_effect}
\end{figure*}

Ten of the $20$ DSHARP sources (noted in Table~\ref{tab:frank_pars}) have \fr residual visibilities that when imaged exhibit a clear two-fold brightness asymmetry in the inner disc, oriented about the disc's major axis. The imaged \fr residuals for these sources are shown in Fig.~\ref{fig:appendix_brightness_asym}. 
{\bf Fig.~\ref{fig:geometric_effect}} demonstrates the most prominent case, Elias~24, in which the asymmetry spans the entirety of the inner disc. This brightness asymmetry across the inner disc can be explained by a geometric viewing effect, provided the disc is optically thick, has finite thickness, and is not viewed exactly face-on. In such a case the observer sees the disc photosphere like the inclined interior of a bowl, where the angle between the local surface normal and the line of sight to the observer varies with azimuth. Since the maximum brightness is seen on the side of the disc surface that is more angled towards the observer (i.e., on the far side of the major axis), the brightness asymmetry can be used to trace the inner disc vertical structure.

This interpretation is supported by considering that among the subsample of $10$ discs in which we see the asymmetry in the \fr imaged residuals, a corresponding asymmetry was identified in the \cl images or their residuals for six sources: in the inner $5 - 10$ au of HD~142666, HD~163296 and Sz~129 \citep{2018ApJ...869L..42H}; and in the core of the survey's three discs with spiral structure, Elias~27, IM~Lup and WaOph~6 \citep{2018ApJ...869L..43H}. The $12$CO $J= 2 - 1$ emission indicates the brighter region is on the disc's far side in all six cases \citep{2018ApJ...869L..42H, 2018ApJ...869L..49I}, consistent with our geometric interpretation.
\citet{2018ApJ...869L..42H} posit the brightness asymmetry in HD~142666, HD~163296 and Sz~129 could be attributed to viewing the interior surface of a finite thickness ring, while we additionally see the asymmetry in sources such as Elias~24, where it spans the entirety of the (fairly smooth) inner disc. \citet{2018ApJ...869L..43H} attribute the brightness asymmetries in the spiral discs to an imperfect determination of the disc phase center, though they note that asymmetric brightness may also be caused by vertical structure. 

Additionally, the $10$ discs in which we see the brightness asymmetry all have a $1.25$ mm optical depth as calculated in \citet{2018ApJ...869L..42H} that is $\approx 1$ in the inner disc (and if the brightness asymmetry is tracing vertical structure, the true optical depth may be $\gg 1$). 
Placing quantitative constraints on vertical scale height and optical depth using the brightness asymmetry will be addressed in a future work.
Investigating potential alternative origins of the observed brightness asymmetry in Appendix~\ref{sec:appendix_brightness_asym}, we find that a simple warp (inclination misalignment between an inner and outer disc) does not yield an asymmetric brightness pattern oriented about the major axis, and an incorrect source phase center does not explain the presence of this asymmetry across so many of the DSHARP sources.

\section{Conclusions}
\label{sec:conclusions}
Finding the effective resolution of \cl images in the DSHARP survey corresponds to an increase in the \cl beam width by an average factor of $1.16$, we used \fr to accurately fit the 1D visibility distribution for each of the $20$ DSHARP sources to a mean factor of $4.3$ longer baseline than brightness profiles extracted from the \cl images and a factor of $3.0$ longer baseline than the \cl models. This yielded super-resolution brightness profiles for each source that more highly resolved azimuthally symmetric (and asymmetric) disc substructure seen in the \cl images. 
The \fr fits additionally identified new features -- an extra gap in the inner $20$ au of SR~4 and Elias~24, as well as new pressure bumps and depressions in the inner $30$ au of HD~142666, HD~163296 and AS~209.
Overall the analysis demonstrated two key points: the DSHARP sources -- already found to ubiquitously contain gaps and rings in \citet{2018ApJ...869L..42H} -- are even more densely structured, especially interior to $30$ au; and the gaps and rings detected in the \cl images, despite in many cases having widths $2 - 3 \times$ that of the \cl beam, become deeper and wider (gaps) or narrower and brighter (rings) when we fit the data with a technique not subject to \cl beam convolution.

We further identified new trends in substructure across the survey:
\begin{itemize}
    \item{substructure in compact discs}: \fr profiles for all three compact ($R_{\rm max} < 50$ au), single-disc systems showed substructure, suggesting it may be frequent in compact sources
    \item{substructure in extended discs}: \fr profiles for six extended ($R_{\rm max} > 50$ au), fairly smooth DSHARP sources found indications of a change in slope in the innermost disc, implying the interior regions of discs may commonly be structured
    \item{potential transition discs}: \fr profiles for two of the oldest discs in the sample suggested they have cleared inner cavities, which may indicate they are dispersing
    \item{spiral arms in disc cores}: \fr profiles for the three single-disc systems with prominent spirals suggested the spiral arms reach into the discs' cores
    \item{inner disc morphologies}: \fr profiles for the three most structured DSHARP discs exhibited highly similar substructure morphology in their inner $40$ au, indicating the same physical processes, e.g., the presence of a companion, may be responsible
    \item{gap morphologies}: \fr profiles for six survey discs that already had prominent gaps in the \cl images showed these features to have greater depth and/or more structure (both within the gap and on its wings)
\end{itemize}

We found that lower values of $\alpha_{\rm turb} / {\rm St}$ than determined in \citet{2018ApJ...869L..46D} and \citet{2020MNRAS.495..173R} are needed to explain the super-resolved ring widths in AS~209, Elias~24, HD~163296, GW~Lup, and HD~143006. 
Finally, the \fr fits also found clear evidence of a geometric viewing effect in $10$ of the $20$ DSHARP sources that traces inner disc vertical structure. 

The extent to which these substructure trends are present in surveys and individual datasets with different biases \citep[DSHARP consists primarily of bright, large discs;][]{Andrews2018} is a question we will address in subsequent work. Those trends that do hold beyond DSHARP may offer the potential to broadly inform open questions on the physical mechanisms underlying dust substructure in protoplanetary discs.

On the technical side, the analysis in this work demonstrated that \fr, and super-resolution fitting techniques more generally, can consistently extract more 1D substructure information from sub-mm disc observations than both \cl images and \cl models. There is a clear limitation with \fr in that it reconstructs the 1D brightness of a source, rather than the 2D brightness as in a \cl image. However, for the purpose of obtaining a 1D brightness profile of a source (under the assumptions of axisymmetry and known source geometry), \fr will yield a more accurate (higher resolution) result, without a loss in sensitivity, compared to extracting an azimuthally averaged profile from the \cl image. Super-resolution techniques can provide new insights from existing datasets, better informing physical inference without requiring deeper and/or longer baseline observations. In practice these tools can also be approachable and efficient; performing a \fr fit requires nontrivial choices for only two hyperparameters (the parameter space for each being small), and the \fr fits shown in this work all took $\lesssim 1$ min to run.
\fr is open source code, available at \href{https://github.com/discsim/frank}{\color{linkcolor}{https://github.com/discsim/frank}} and documented at \href{https://discsim.github.io/frank}{\color{linkcolor}{https://discsim.github.io/frank}}. All \fr fits in this work are available at \href{https://zenodo.org/record/5587841}{\color{linkcolor}{https://zenodo.org/record/5587841}}.

\section*{Acknowledgements}
JJ thanks H.~Vaivao for his comments on the work. This work was supported by the STFC consolidated grant ST/S000623/1. G.R. acknowledges support from the Netherlands Organisation for Scientific Research (NWO, program number 016.Veni.192.233) and from an STFC Ernest Rutherford Fellowship (grant number ST/T003855/1). This work has also been supported by the European Union's Horizon 2020 research and innovation programme under the Marie Sklodowska-Curie grant agreement No. 823823 (DUSTBUSTERS). 

{\it Software:} 
\texttt{NumPy} \citep{doi:10.1109/MCSE.2011.37}, 
\texttt{SciPy} \citep{virtanen2019scipy},
\texttt{Matplotlib} \citep{doi:10.1109/MCSE.2007.55},
\texttt{Astropy} \citep{astropy:2013, astropy:2018},
\texttt{Jupyter Notebook} \citep{Kluyver:2016aa},
\texttt{CASA} \citep{2007ASPC..376..127M}, 
\texttt{uvplot} \citep{uvplot_mtazzari}

\section{Data availability}
The data underlying this article are available in the DSHARP Data Release at \href{https://bulk.cv.nrao.edu/almadata/lp/DSHARP}{https://bulk.cv.nrao.edu/almadata/lp/DSHARP}. The datasets were derived from this previously released source in the public domain.

\bibliographystyle{mnras}
\bibliography{references.bib}

\appendix

\section{Point source-corrected fits}
\label{sec:appendix_point_source}
\begin{figure*}
	\includegraphics[width=\textwidth]{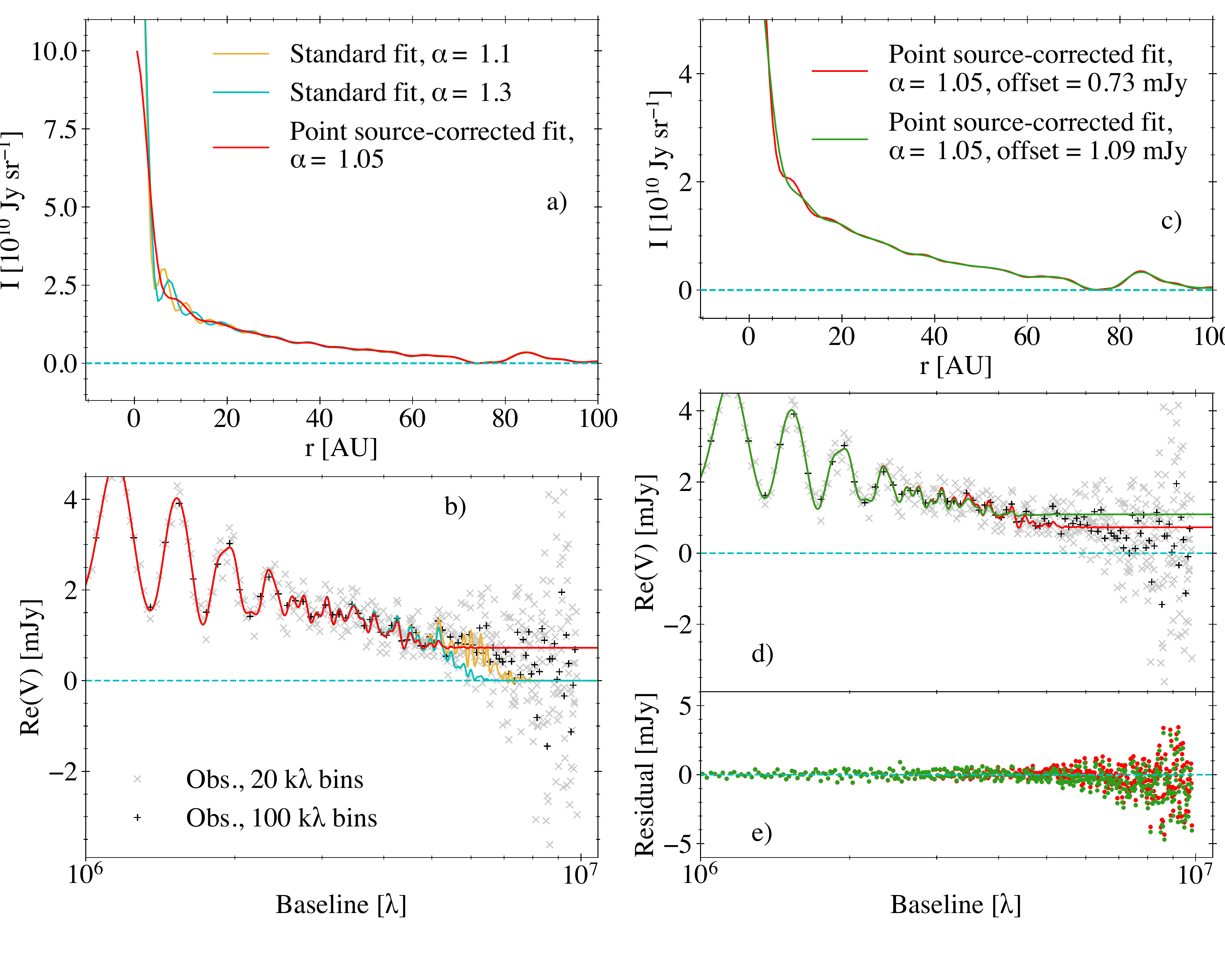}
	    \caption{{\bf Effects of a point source-corrected fit} \newline 
	    a) \fr brightness profiles for two standard fits using different \al, and the profile for the point source-corrected fit shown in the main text. \newline
	    b) A zoom on the data's long baselines ($> 1.0$ \ml, corresponding to spatial scales $< 0.2 \arcsec$; data shown in $20$ and $100$ \kl bins), the two standard \fr fits, and the point source-corrected fit. \newline 
	    c) \fr brightness profiles for the point source-corrected fit in (a), and a point source-corrected fit using a $1.5\times$ larger point source amplitude. \newline 
	    d) As in (b), but for the two point source-corrected fits in (c). \newline 
	    e) Residuals (in $20$ \kl bins) of the two point source-corrected fits in (d).
        }
    \label{fig:appendix_point_source}
\end{figure*}

To demonstrate the effect of a point source-corrected fit, {\bf Fig.~\ref{fig:appendix_point_source}}(a) -- (b) compares a model generated with this approach to two standard \fr fits for GW~Lup. In panel (b), the observed visibilities remain systematically positive at the longest baselines, i.e., do not converge on zero. Their offset is $0.7$ mJy; for reference, Re(V) plateaus at $88.9$ mJy at short baselines. First considering the two standard \fr fits (which use different \al values), the model with \al $= 1.1$ fits the visibilities out to $\approx 7$ \ml, at which point some of the $100$ \kl binned values approach zero. However because the data are noise-dominated by this baseline, the corresponding brightness profile in Fig.~\ref{fig:appendix_point_source}(a) has noisy oscillations, most apparent at small radii. By comparison, increasing \al to $1.3$ effectively fits the data to shorter baseline, $\approx 5$ \ml, beyond which the binned SNR start to dither about SNR $=1$. The model drives toward zero (by design) once its SNR threshold is reached, which is problematic if the fit's slope at this baseline is steeper than the average slope of the true, underlying signal in the data. That appears to be the case here, as the fit's slope still translates to strong oscillations in the brightness profile in panel (a).

The point source-corrected model in Fig.~\ref{fig:appendix_point_source}(b) fits the data out to comparable baseline to the \al $= 1.3$ case, but once its SNR threshold is reached, the fit takes on a constant visibility amplitude (rather than driving toward zero). This amplitude is the mean of the data beyond the baseline at which the $20$ \kl binned SNR first drops below unity. The strong oscillations in the innermost disc present in the standard fits are no longer apparent in the point source-corrected fit, though we do still see some small amplitude oscillations across all radii in the brightness profile, whose sensitivity we will examine below. 
The fit's zero slope over the data's longest baselines yields a conservative representation of features on the corresponding spatial scales in the brightness profile, which we prefer because of the ambiguity in where the true visibility signal converges on zero. 

While for practical purposes the point source-corrected model is the best approach we have at present to fit a visibility distribution that does not clearly converge on zero, it has limitations. First, because it involves fitting \fr to a visibility distribution from which we have subtracted a constant offset, the SNR of the resulting data are not identical to those of the observed data. This is why the point source-corrected model in Fig.~\ref{fig:appendix_point_source}(b) fits the visibilities beyond $\approx 4$ \ml less closely than the shown standard fits, despite using a lower \al. 

Second, while we have determined the point source amplitude by taking the mean of the longest baseline visibilities, they are in general dominated by noise and so not necessarily an accurate indication of the true signal. We thus test how the applied point source offset affects the \fr visibility fit and in turn substructure in the brightness profile. Fig.~\ref{fig:appendix_point_source}(d) shows the visibility fit for GW~Lup when we increase the point source offset to $1.5\times$ the mean of the long baseline data. This offset expectedly yields larger amplitude (negative) residuals in panel (e), while also reducing structure in the brightness profile interior to $\approx 0.1 \arcsec$ in panel (c). The reduced prominence of structure seems less correct than the fit with a lower point source offset based on the residuals in (e). However it is also not clear that the structure interior to $0.1 \arcsec$ in the smaller point source offset fit is real; this ambiguity motivates our treatment of the difference between these two fits as an informal uncertainty estimate in all discs where we use the point source-corrected model in the main text.

\section{Residual image brightness asymmetries}
\label{sec:appendix_brightness_asym}
\begin{figure*}
	\includegraphics[width=\textwidth]{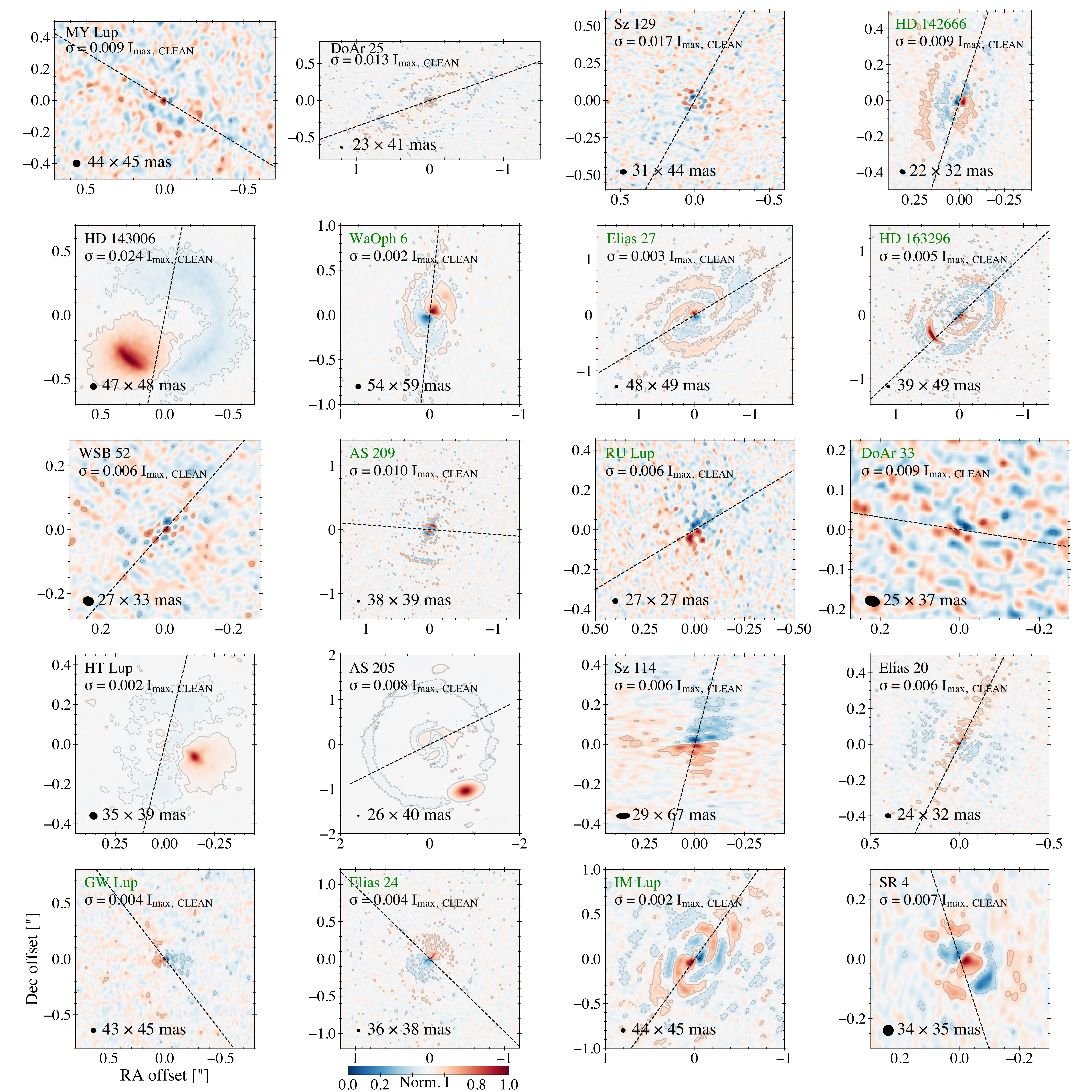}
	    \caption{{\bf \fr imaged residuals} \newline 
	    The \fr residual visibilities imaged (0 \cl iterations), with $\pm3 \sigma$ contours overplotted ($\sigma$ is given for each image), and a dashed line along the fitted position angle. The residual image is convolved with the published \cl beam and uses a linear color scale. Discs are ordered as in Fig.~\ref{fig:profiles_grid}. The $10$ sources that exhibit a clear two-fold brightness asymmetry in the inner disc have their names shown in green. All images use a linear color scale (a normalized color bar is shown, and the $\sigma$ value for each image is given).
        }
    \label{fig:appendix_brightness_asym}
\end{figure*}

Considering the residual brightness asymmetries in Sec.~\ref{sec:geometric_effect}, {\bf Fig.~\ref{fig:appendix_brightness_asym}} shows the \fr residuals imaged for each DSHARP source.
Here we present tests to determine whether the observed trend of a brightness asymmetry oriented about the major axis in $10$ of the $20$ sources could -- instead of a geometric effect -- be produced by either an incorrect source phase center or a simple warp in the form of a misalignment between the inner and outer discs (effectively an incorrect inclination). First considering a phase center error, shifting the phase center of a flat disc generates an asymmetry in the direction of the centroid error. In order to explain the observed asymmetry pattern in $10$ of the $20$ DSHARP discs would thus require that some aspect of fitting for the phase center (which was done by fitting a 2D Gaussian to the image) biased the error toward alignment with the disc's minor axis. We do not see how such a bias could arise.

Nevertheless, as a precaution we considered the
$1\sigma$ uncertainties in fitted right ascension and declination offsets as determined in \citet{2018ApJ...869L..42H}, which are typically $1 - 3$ mas. To test whether shifting the phase center within this range could effectively erase the brightness asymmetry in the residual maps, for each DSHARP source we applied a phase center that differed from the published value by $1$ or $3$ mas, with the perturbation oriented along the disc's minor axis as well as at $\pi / 4$ intervals over the full $2\pi$ in azimuth. For each of these applied phase centers, we then fit for the \fr profile, and compared the resulting imaged \fr residuals. Shifting the phase center in this way did change the amplitude of the brightness asymmetry in the inner disc by a factor of $\lesssim 2$, and in some cases it slightly rotated the asymmetry's orientation. But in almost all cases the asymmetry clearly persisted, suggesting it is not an artifact of an incorrect phase center.

For the $10$ DSHARP discs in which we initially did not identify a clear brightness asymmetry, shifting the phase center along the disc's minor axis could in some cases create an asymmetry similar to that observed. The same was true for mock datasets in which we intentionally assigned an incorrect phase center. 
And $2$ of these $10$ sources, SR~4 and Sz~114, exhibited an asymmetry that was not aligned about the major axis; however shifting the phase center within published uncertainty ($<3$ mas) could reorient the asymmetry about the major axis.
Taking all of this together, again we do not see why fitting for the phase center as described in \citet{2018ApJ...869L..42H} would introduce a bias along the disc's minor axis.

Next considering disc misalignment, we forward modeled mock observations emulating DSHARP datasets that have an inner disc separated from an outer ring by a deep gap. We generated images in which the inner disc's inclination was misaligned relative to the outer ring by values between $0.1 - 3^{\rm o}$ (the published $1\sigma$ uncertainties on inclination are $\leq 2^{\rm o}$ in either direction). We then forced the geometry used to deproject the source to be that of the outer ring (separately, we also ran trials in which we fit for the geometry using a 2D Gaussian in visibility space), and fit the deprojected dataset with \fr. We found that a misaligned inner disc produces a {\it four-fold symmetric} pattern oriented equivalently about the major or minor axis in the imaged \fr residuals. In the real observations we instead see a {\it two-fold asymmetric} pattern oriented about the major axis.
 
\bsp
\label{lastpage}
\end{document}